\documentclass[12pt]{iopart}
  
\usepackage{mciteplus}
\usepackage{graphicx,color}
\usepackage{amssymb}
\usepackage{subfigure}

\def\beq{\begin{equation}}
\def\eeq{\end{equation}}
\def\bea{\begin{eqnarray}}
\def\eea{\end{eqnarray}}

\def\tr{\textcolor{red}}
\def\tb{\textcolor{blue}}

\begin{document}

\title[Collective force generated by multiple biofilaments]{Collective force generated by multiple biofilaments can exceed
the sum of forces due to individual ones}

\author{Dipjyoti Das}
\address{Department of Physics, Indian Institute of Technology, Bombay,
Powai, Mumbai-400 076, India}
\ead{dipjyoti@phy.iitb.ac.in}
\author{Dibyendu Das}
\address{Department of Physics, Indian Institute of Technology,
Bombay, Powai, Mumbai-400 076, India}
\ead{dibyendu@phy.iitb.ac.in}

\author{Ranjith Padinhateeri}
\address{Department of Biosciences and Bioengineering, Indian Institute of Technology, Bombay, Powai, Mumbai-400 076, India}
\ead{ranjithp@iitb.ac.in}

\date{\today}

\begin{abstract}
Collective dynamics and force generation by cytoskeletal filaments are crucial in many cellular processes. Investigating 
growth dynamics of a bundle of $N$ independent cytoskeletal filaments pushing against a wall, we show that  chemical switching (ATP/GTP hydrolysis) leads to a collective phenomenon that is currently unknown. Obtaining force-velocity relations for different models that capture chemical switching, we show, analytically and numerically, that the collective stall force of $N$
filaments is {\it greater} than $N$ times the stall force of a single
filament. Employing an exactly solvable toy model, we analytically prove the above result for $N=2$. We, further, numerically show the existence of this collective phenomenon, for $N\ge2$, in realistic models (with random and sequential hydrolysis) that simulate actin and microtubule bundle growth. We make quantitative predictions for the excess forces, and argue that this collective effect is related to the {\it non-equilibrium} nature of chemical switching.

\end{abstract}



\submitto{\NJP}
\maketitle


\section{Introduction}

Biofilaments such as actin and microtubules are simple nano-machines
that consume chemical energy, grow and generate significant amount of force.
Living cells make use of this force in a number of ways -- e.g., to generate protrusions
and locomotion, to segregate chromosomes during cell devision, and to perform specific tasks such as acrosomal process 
where an actin bundle from sperm penetrates into an egg~\cite{howard-book,Alberts-book}. Understanding the growth of cytoskeletal filaments provides insights into a wide range of questions related to
collective dynamics of biomolecules and chemo-mechanical energy transduction.

Actin and microtubules  grow by addition of subunits that are typically bound to ATP/GTP. Once polymerised, ATP/GTP in these subunits  get hydrolysed into ADP/GDP creating a heterogeneous filament with interesting polymerisation-depolymerization dynamics~\cite{Pollard1986,Desai_Mitchison_MT:97}. {In actin, the subunits are known to also exist in an intermediate state bound to ADP-${\rm P_i}$~\cite{Korn-Calier:87,vavylonis:05,Jegou2011}. Even though the growth kinetics of actin and microtubules are similar, under cellular conditions they show diverse dynamical phenomena such as treadmilling and dynamic instability. They also have very different structures: actin filaments are two-stranded helical polymers while microtubules are hollow cylinders made of 13 protofilaments~\cite{howard-book,Alberts-book, Desai_Mitchison_MT:97}.} 

These filaments  growing against a wall can generate force using a Brownian ratchet mechanism~\cite{howard-book}.
The maximum force these filaments can generate, known as ``stall force", is of great interest to experimentalists and theorists alike~\cite{howard-book,footer-dogterom:07,lacoste11}. 
In theoretical literature, growth of a single bio-filament and the resulting force generation
 has been extensively studied~\cite{howard-book,hill:81, kolomeisky:06, vavylonis:05}. 
  Explicit relations of velocity versus force (or monomer concentration)  have been
derived assuming either
simple polymerization and depolymerization rates~\cite{howard-book} or more realistic
models that take into account ATP/GTP hydrolysis~\cite{hill:85,Leibler-cap:96,kolomeisky:06, vavylonis:05,Ranjith2009,Ranjith2012,Manoj:2013}. In fact without considering hydrolysis, the observed velocities and length fluctuations of single filaments cannot be explained~\cite{hill:85,Leibler-cap:96,kolomeisky:06, vavylonis:05}.

Even though single filament studies teach us useful aspects of kinetics of the system, what
is relevant, biologically, is the collective behaviour of multiple filaments.
However, the theory of collective effects due to $N$
($\geq 2$) filaments pushing against a wall is  poorly understood.  Simple models of filaments with polymerization and
depolymerization rates have been studied: for two filaments, exact
dependence of velocity on force is known \cite{lacoste11,
  kolomeisky:05}, while for $N \geq 2$, numerical results and theoretical arguments \cite{van-doorn:00,lacoste11}
  show that the net force $f_s^{(N)}$ to stall the system is $N$ times the force $f_s^{(1)}$ to stall a single filament. 
A similar result, namely $f_s^{(N)} \propto N$, was obtained for multiple protofilaments with lateral interactions in the absence of hydrolysis \cite{Kierfeld:2011}. Under harmonic force, experimental studies claimed collective stall forces to be lesser than $N$ times single filament stall force for actin \cite{footer-dogterom:07}, and proportional to $N$ for microtubules \cite{Laan-pnas:08}. 
 Based on the understanding of single filaments, it has been speculated that hydrolysis might lower the stall force of $N$ actin filaments~\cite{footer-dogterom:07}. In a recent theoretical study within a two-state model, under harmonic force \cite{Kierfeld:2013}, it was shown that the average polymerization  forces generated by $N$ microtubules scale as $N$.
However, for constant force ensemble, there exists no similar theory for multiple filaments that incorporates the crucial feature of ATP/GTP hydrolysis, or appropriate structural transitions~\cite{mitchison:2009}, systematically. How, precisely, the ATP/GTP hydrolysis  influences the growth and force generation of N filaments is an important open question.

Motivated by the above, in this
paper, we investigate collective dynamics of multiple filaments, using {a number of} models  that capture chemical switching.
{These models extend the work of Tsekouras et al.~\cite{lacoste11} by adding the ``active" phenomenon of ATP/GTP hydrolysis.}
The first one is a simple model in which each filament
switches between two depolymerization states. Within the model, we
analytically show that $f_s^{(2)} > 2 f_s^{(1)}$; 
this result extends to $N>2$.  We then proceed to 
study numerically two detailed models involving  \emph{sequential} and \emph{random} mechanisms for hydrolysis~\cite{kolomeisky:06, vavylonis:05, Ranjith2009,Ranjith2012}.
 Using parameters appropriate for the 
cytoskeletal filaments, we show that {indeed $f_s^{(N)} > N f_s^{(1)}$. The excess force ($\Delta^{(N)}=f_s^{(N)} - N f_s^{(1)}$) being $\sim 1-9$ pN for microtubules,
and $\sim 0.1-1.5$pN for actin, is detectable in appropriately designed experiments.
Finally we show the robustness of our results by considering realistic variants of the detailed models.  }


\section{Models and Results}

\subsection{An exactly solvable toy model that demonstrates the relationship $f_s^{(2)} > 2f_s^{(1)}$ analytically}
\label{toy}

\begin{figure}
\centering
\includegraphics[scale = 0.27]{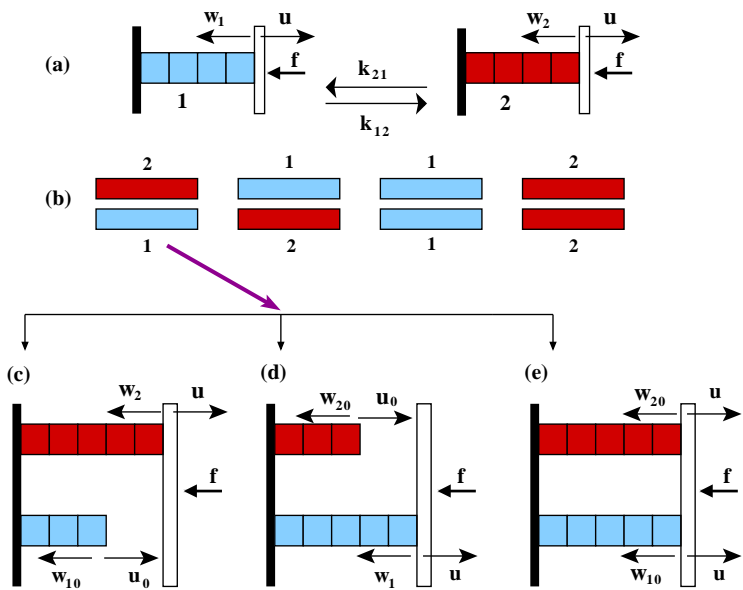}
\caption{Schematic depiction of the toy model: (a) Single filament dynamics (b) Four possible
states of the two-filament system. (c-e) Microscopic dynamics of the filaments in the \{2,1\} state. In all the cases, the left wall is fixed, while the right wall is movable against a resistive force $f$.}

\label{model}
\end{figure}
{We first discuss a simple toy model which is analytically tractable and hence demonstrates the phenomenon exactly.} 
We consider $N$ filaments, {each composed of subunits of length $d$}, 
collectively pushing a rigid
wall, with an external force $f$  acting against the growth
direction of the filaments (see Fig.~\ref{model}). Consistent with Kramers theory, each filament has a growth rate $u = u_0{\rm e}^{-\tilde{f}\delta}$ 
 when they touch the wall, and $u_0$
otherwise; {here $\tilde{f}=fd/k_{\rm B}T$ and $\delta \in [0,1]$ is the force distribution factor~\cite{lacoste11,Dogterom-yurke:97}.} 
Each filament can be in two
distinct depolymerization states $1$ (blue) or $2$ (red)  (Fig.~\ref{model}a) giving rise to four distinct states for a two-filament system (Fig.~\ref{model}b). Filaments in state $1$ and $2$ depolymerise with intrinsic rates $w_{10}$ and $w_{20}$ respectively. When only one filament belonging to a multifilament system is in contact with the wall, these rates become force-dependent, and is given by $w_1 = w_{10} {\rm e}^{\tilde{f}(1-\delta)}$ or $w_2 = w_{20} {\rm e}^{\tilde{f}(1-\delta)}$ (Fig.~\ref{model}c,d). When more than one filament touch the wall simultaneously (Fig.~\ref{model}e), depolymerisation rates are force independent (similar to ref.~\cite{lacoste11}) as a single depolymerisation event does not cause wall movement
{for perfectly rigid wall and filaments.}
 Furthermore, any filament as a whole
can dynamically switch from state $1$ to $2$ with rate $k_{12}$, and
switch back with rate $k_{21}$ (see Fig.~\ref{model}a). 
{Below we focus on $\delta = 1$, consistent with earlier theoretical
literature \cite{Ranjith2009,lacoste11,kolomeisky:05} and close to experiments on microtubules \cite{Dogterom-yurke:97}.}

For a single filament,  the average velocity is given by $V^{(1)} = (u - w_1) P_1 + (u - w_2) P_2$,
where $P_1$ and $P_2$ denote the probability 
of residency in states $1$ and $2$. 
Following Fig.~\ref{model}a, or using Master equations
for the microscopic dynamics (see  \ref{toy-1filament})
it can be shown that the detailed balance condition $P_1 k_{12} = P_2 k_{21}$
holds in the steady state. Along with 
the normalization condition $P_1 + P_2 = 1$, this yields:
\beq
 V^{(1)} = {[(u-w_1) k_{21} + (u - w_2) k_{12}]}/{(k_{12} +
    k_{21})}. 
\label{V1} 
\eeq 

Setting $V^{(1)} = 0$, the stall force of the single filament is
$f_s^{(1)} = \frac{k_BT}{d}\ln [(k_{12}+k_{21})u_0/(k_{12} w_{20} + k_{21}
  w_{10})]$, 
which is independent of the value of $\delta$.

For two filaments, let $P_{11}$, $P_{12}$, $P_{21}$, and $P_{22}$
denote the joint probabilities for filaments to be in states
$\{1,1\}$,$\{1,2\}$, $\{2,1\}$, and $\{2,2\}$, respectively (Fig.~\ref{model}b).
Using the microscopic Master equations (see \ref{toy-2filament-bal-eq}  and Fig. \ref{states}) the following 
steady state balance conditions may be derived: 
$k_{21} (P_{12} + P_{21}) = 2 k_{12} P_{11}$, $k_{12}
(P_{12} + P_{21}) = 2 k_{21} P_{22}$, and $k_{12} P_{11} + k_{21}
P_{22} = (k_{12} + k_{21}) P_{12}$. The latter conditions, in addition
to the normalization condition $\sum_{i,j} P_{ij} =1$, solve for  $P_{11} =
k_{21}^2/(k_{12} + k_{21})^2$, $P_{22} = k_{12}^2/(k_{12} + k_{21})^2$
and $P_{12} = P_{21} = k_{12}k_{21}/(k_{12} + k_{21})^2$. The
average velocity of the wall  for the two-filament system is $V^{(2)}
= P_{11} v_{11} + P_{12} v_{12} + P_{21} v_{21} + P_{22} v_{22}$.
Here $v_{11}$ and $v_{22}$ are the velocities in homogeneous states
$\{1,1\}$ and $\{2,2\}$ respectively. These  
are known from previous works
\cite{kolomeisky:05,lacoste11}:
\begin{eqnarray}
v_{11} &=& 2 (u u_0 - w_1 w_{10})/(u + u_0 + w_1 + w_{10}),\nonumber\\
v_{22} &=& 2 (u u_0 - w_2 w_{20})/(u + u_0 + w_2 + w_{20}),
\label{v1122}
\end{eqnarray} 
where $i=1,2$. The velocities in the heterogeneous states ($\{1,2\}$ or 
$\{2,1\}$) have not been calculated previously; solving the inter-filament gap dynamics (see \ref{toy-2filament-v12})
we get
\begin{equation}
v_{12} = v_{21} = \frac{2u - (u+w_1)\gamma_1 - (u+w_2)\gamma_2 + (w_1+w_2)\gamma_1\gamma_2}{(1 - \gamma_1 \gamma_2)}.
\label{v12}
\end{equation} 
Here $\gamma_1 =(u + w_{20})/(u_0 + w_1)$ and $\gamma_2 =(u +
w_{10})/(u_0 + w_2)$ are both $< 1$ for the existence of the
steady state. Combining all these, we find that the velocity of the two-filament system, switching between two states, is
\beq
V^{(2)} = {[k_{21}^2 v_{11} + k_{12}^2 v_{22} 
+ 2 k_{12} k_{21} v_{12}]}/{(k_{12} + k_{21})^2} 
\label{V2}
\eeq
with  $v_{11}$, $v_{22}$, $v_{12}$ and $v_{21}$
given by Eqs. (\ref{v1122}) and (\ref{v12}). Note that various limits 
($w_{10} = w_{20}$, $k_{12}=0$, or $k_{21} \rightarrow \infty$) retrieve
the expected result $V^{(2)} = v_{11}$.

\begin{figure}
\centering
\includegraphics[scale=0.55]{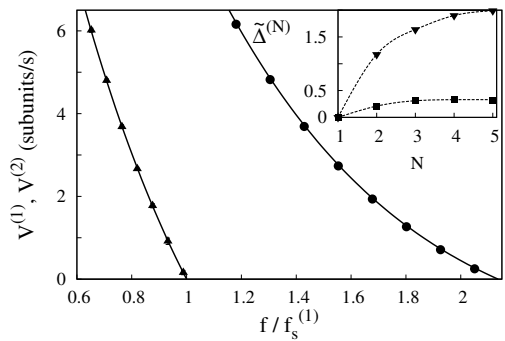}
\caption{ Main figure: Scaled force-velocity relation for one ($V^{(1)}$,
  $\blacktriangle$) and two ($V^{(2)}$, $\bullet$) filaments with
  switching. The continuous curves are analytical formulae given
  by Eqs. (\ref{V1}) and (\ref{V2}). The data points 
($\blacktriangle$ and $\bullet$) are from kinetic 
  Monte-Carlo simulations with $\delta = 1$.  Inset: Excess stall force $\tilde{\Delta}^{(N)}$, with varying number of filaments ($N$), at $\delta = 1$ ($\blacksquare$) and $\delta=0$ ($\blacktriangledown$).   All results are for parameters $u_0=40s^{-1}$, $w_{10}=1s^{-1}$,$w_{20}=15s^{-1}$, and $k_{12}=k_{21}=0.5s^{-1}$.}
\label{V-f}
\end{figure}

The Eq. (\ref{V2}) is valid for any $\delta$. To obtain stall force we
set $V^{(2)}  = 0$ and this leads to a cubic equation (for $\delta=1$) in ${\rm e}^{\tilde{f}}$ whose only
real root gives $f_s^{(2)}$ analytically (see \ref{toy-2filament-fs2}). 
The analytical result for $V^{(1)}$ and
$V^{(2)}$ (Eqs. (\ref{V1}) and (\ref{V2})) are plotted in
Fig. \ref{V-f} (main figure) as continuous curves, and the data points obtained from
kinetic Monte-Carlo simulations, with the same
parameters, are superposed on them.  Most importantly we see 
 that the scaled force $f_s^{(2)}/f_s^{(1)}$ for which $V^{(2)}
= 0$ is clearly $> 2$.
For $N > 2$ filaments, we do not have any analytical formula for the
model, but we obtain stall forces from the kinetic Monte-Carlo
simulation. We plot the excess force $\tilde{\Delta}^{(N)} = \tilde{f}_s^{(N)} - N \tilde{f}_s^{(1)}$
for different $N$ in the inset of Fig. \ref{V-f}, {for $\delta = 1$ and $\delta=0$. As can be seen, 
$\tilde{\Delta}^{(N)} > 0$ increases with $N$, and seems to saturate at large
$N$ --- this is true for all $\delta \in [0,1]$.} 
Thus we have shown that dynamic switching between heterogeneous
depolymerization states lead to $f_s^{(N)} \neq N f_s^{(1)}$.

\begin{figure}
\centering
\includegraphics[scale=0.5]{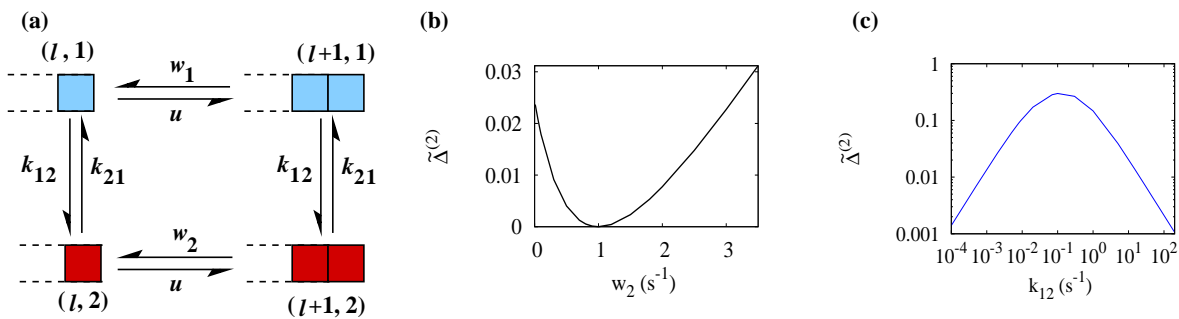}
\caption{(a) Schematic depiction of a connected loop in the configuration  space of single-filament toy model. The configurations are denoted by ordered pairs, whose first element is the instantaneous length and second element is the depolymerization state ($1$ or $2$).  (b) Excess  force $\tilde{\Delta}^{(2)}$  versus $w_{2}$ (with $w_{1} = 1s^{-1}$). The parameters are $u_0 = 40s^{-1}$, and  $k_{12} = k_{21} = 0.5s^{-1}$. (c) $\tilde{\Delta}^{(2)}$  versus $k_{12}$ (with $k_{21} = 0.5s^{-1}$).  Other parameters are $u_0 = 40s^{-1}$,  $w_{1} = 1s^{-1}$, and $w_{2} = 15s^{-1}$. In both Figs (b) and (c), $\delta=1$ (i.e $w_1=w_{10}$ and $w_2=w_{20}$).}
\label{fig3}
\end{figure}

{ We now proceed to show that the introduction of switching between distinct  chemical states ($w_1 \neq w_2$, $k_{12}/k_{21} \neq 0$ or $k_{12}/k_{21} \neq \infty$) produces non-equilibrium  dynamics embodied in the violation of the well known {\it detailed balance} condition. To demonstrate this for the single-filament toy model, in Fig. \ref{fig3}a, we consider a loop of dynamically connected configurations (charaterised by its length and state):  $\{l,1\} \rightleftharpoons \{l+1,1\} \rightleftharpoons \{l+1,2\}  \rightleftharpoons \{l,2\} \rightleftharpoons \{l,1\}$. The product of rates clockwise  and anticlockwise  are $u k_{12} w_2 k_{21}$ and $k_{12} u k_{21} w_1$, respectively. For the condition of detailed balance to hold (in equilibrium) the two products need to be equal (Kolmogorov's criterion), which leads to $w_1 = w_2$. In Fig. \ref{fig3}b, we plot $\tilde{\Delta}^{(2)} = \tilde{f}_s^{(2)} - 2\tilde{f}_s^{(1)}$ against $w_{2}$ (with fixed $w_{1} = 1s^{-1}$) --- we see that $\tilde{\Delta}^{(2)}> 0$ for all $w_{2}$ except $w_{2} = w_{1}$ (the equilibrium case). This indicates that the collective phenomenon of excess force generation is tied to the departure from equilibrium. This should be compared with Ref. \cite{van-doorn:00}, where it was shown that for a  biofilament model involving no switching (which is unrealistic), the relationship $f_s^{(N)} = N f_s^{(1)}$ follows from the condition of detailed balance.  The effect of non-equilibrium switching is further reflected  in the variation of $\tilde{\Delta}^{(2)}$ as a function of switching rates. If  $k_{12}$ is varied (keeping  $k_{21}$ fixed), we see in Fig. \ref{fig3}c that $\tilde{\Delta}^{(2)} > 0$ always, except in the limits $k_{12}/k_{21} \rightarrow 0$ or $\infty$. These limits correspond to  the absence of switching and hence equilibrium.}

Is our toy  model comparable to real cytoskeletal filaments with ATP/GTP hydrolysis?
{In real filaments the tip monomer can be in two states -- ATP/GTP bound or ADP/GDP bound -- similar to states $1$ or $2$ of our toy model.} 
{However, in real filaments the chemical states of the subunits may vary along the length}, and the switching probabilities are indirectly coupled to force-dependent polymerisation and depolymerisation events~\cite{hill:85,Leibler-cap:96,kolomeisky:06, vavylonis:05,Ranjith2009,Ranjith2012,sumedha}.
{ Thus study of more complex models with explicit ATP/GTP hydrolysis are warranted to get convinced that the interesting collective phenomenon is expected in real biofilament experiments {\it in vitro}.}

\subsection{Realistic models with random hydrolysis confirm the relationship  $f_s^{(N)} > Nf_s^{(1)}$ }

 In the literature, there are  three different  models 
of  ATP/GTP hydrolysis, namely the sequential hydrolysis model~\cite{kolomeisky:06,Ranjith2009} and the random hydrolysis model~\cite{Ranjith2012,vavylonis:05, Antal-etal-PRE:07}, and a mixed cooperative hydrolysis model~\cite{Leibler-cap:96, Li:2009, Kolomeisky:2013}.
 In this section, we investigate the collective dynamics within the random model as it is a widely used model and is thought to be closer to reality~\cite{Jegou2011}. We also present  different variants of the random model to show that our results are robust. In  \ref{sequential}, interested readers may find similar results (with no qualitative difference) for the sequential hydrolysis model.

\begin{figure}
\centering
\includegraphics[scale=0.42]{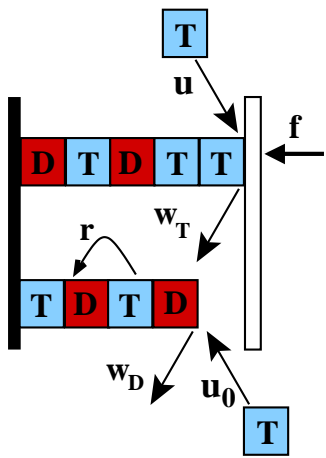}
\caption{Schematic diagram of two-filament random hydrolysis model. ATP/GTP and ADP/GDP subunits are shown as letters 'T' (blue) and 'D' (red)  respectively. Here the switching ATP/GTP $\rightarrow$ ADP/GDP can happen randomly at any ATP/GTP subunit. Various processes are shown in arrows and by corresponding rates, as discussed in the text.}
\label{random-hydro}
\end{figure}

In Fig. \ref{random-hydro} we show the schematic diagram of the random hydrolysis model.  In this model, polymerization of
filaments occurs with a rate $u = u_0 {\rm e}^{-\tilde{f}}$ (next to the wall) or
$u_0 $ (away from the wall). Note that $u_0=k_0c$ where $k_0$ is the intrinsic polymerization rate constant and $c$ is the free ATP/GTP subunit concentration. The depolymerization occurs with a rate  
$w_T$ if the tip-monomer is ATP/GTP bound, and $w_D$ if it is
ADP/GDP bound. There is no $f$ dependence of $w_T$ and $w_D$ (i.e. $\delta = 1$).  In the random model, hydrolysis happens on any subunit
randomly in space \cite{Ranjith2012} (see Fig. \ref{random-hydro}) with a rate $r$ per unit ATP/GTP bound monomer. { Here, as argued in ref~\cite{Ranjith2009,Ranjith2010}, we consider effective lengths of a tubulin and G-actin subunits as $0.6$nm and $2.7$nm respectively to account for the actual multi-protofilament nature of the real biofilaments (see \ref{1layer} for details).}
We did kinetic Monte-carlo simulations of this model using realistic parameters suited for microtubule and actin  (see Table~\ref{table1}).

%
\begin{table}
\caption{Rates ( Actin~\cite{Pollard1986, howard-book} and Microtubules (MT) ~\cite{Desai_Mitchison_MT:97, howard-book, mitchison:1984})}
\label{table1}
\begin{indented}
\item[]
\begin{tabular}{|c|c|c|c|c|} \hline
~ & $k_0$ ($\mu M^{-1}s^{-1}$) & $w_T ~(s^{-1}) $ & $w_D ~(s^{-1}) $  & $r$ ($s^{-1}$) \\ \hline
Actin & $11.6$ & $1.4$ & $7.2$  & $0.003$ \\ \hline
MT & $3.2$ & $24$ & $290$  & $0.2$ \\ \hline
\end{tabular}
\end{indented}
\end{table}

\begin{figure}
\centering
\includegraphics[scale = 0.55]{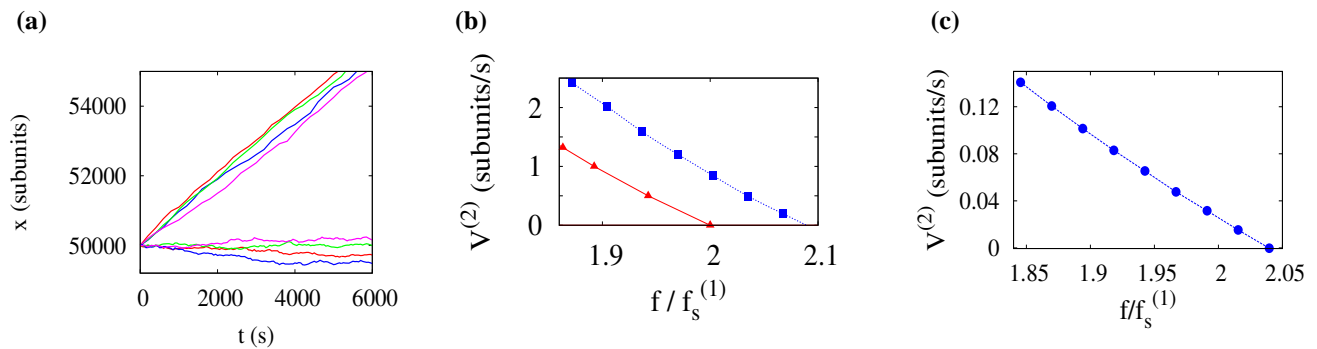}
\caption{ Results for two-filaments with random hydrolysis (see Table~\ref{table1} for parameters): (a) Different traces of wall position ($x$) vs. time ($t$)  of two-microtubule system, at $f = 2f_s^{(1)}$ (top), and  at the stall force $f = f_s^{(2)}$ (bottom).   (b) Scaled force-velocity relations for two microtubules  with  random hydrolysis (\tb{$\blacksquare$}), and no hydrolysis (\tr{$\blacktriangle$}). (c) Scaled force-velocity relations for two actins with  random hydrolysis. Concentrations are $c=100\mu$M for microtubules and $c=1\mu$M for actins. } 
\label{fig4}
\end{figure}

{We first numerically calculated the single microtubule stall force $f_s^{(1)}$, and then we checked that for a two-microtubule  
system; the wall moves with a positive velocity at $f = 2 f_s^{(1)}$ (see Fig.~\ref{fig4}a (top)). We find the actual 
stall force  $f_s^{(2)}$ at which the wall has zero average velocity (Fig.\ref{fig4}a (bottom)) is  greater than $2 f_s^{(1)}$. In Fig.~\ref{fig4}b(\tb{$\blacksquare$}) the velocity against scaled force 
for two microtubules  show clearly that $f_s^{(2)} > 2 f_s^{(1)}$, 
and the resulting  $\Delta^{(2)}= 0.09 \times f_s^{(1)}=1.51$pN (where $f_s^{(1)}=16.75$pN for $c=100\mu$M). Another interesting point is that at any velocity, even away from stall,  the collective force with hydrolysis is way above the collective force without hydrolysis ---  a comparison of the force-velocity curves without 
hydrolysis ($r=0$; Fig.~\ref{fig4}b, \tr{$\blacktriangle$}) and with random hydrolysis (Fig.~\ref{fig4}b, \tb{$\blacksquare$}) demonstrate this.  Similar  force-velocity curve for two-actins is shown in Fig.~\ref{fig4}c and we calculated $\Delta^{(2)}= 0.038 \times f_s^{(1)}=0.119$pN (where $f_s^{(1)}=3.134$pN for $c=1\mu$M).
In Fig.~\ref{excess_force}a and \ref{excess_force}b, we show that the excess stall force $\Delta^{(N)}$ increases with $N$, both for microtubule and actin. For microtubule, the excess force is as big as $6.5$ pN for $N=8$, while for actin it goes up to $0.6$ pN for $N=8$. 

\begin{figure}
\centering
\includegraphics[scale = 0.55]{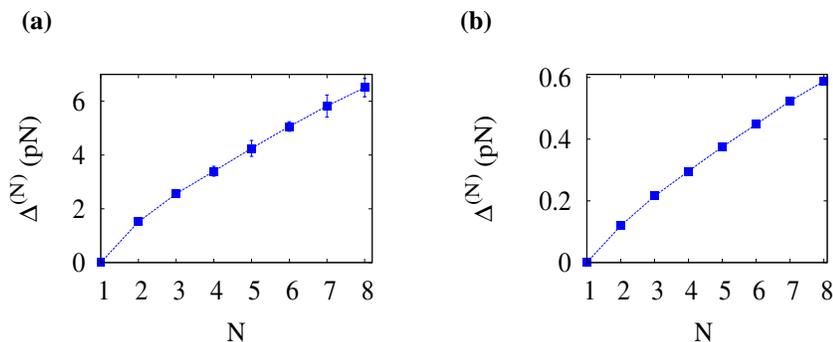}
\caption{ Excess stall force $\Delta^{(N)}$ against filament number $N$, with random hydrolysis for (a) microtubules and (b) actins. Concentrations are $c=100\mu$M for microtubules and $c=1\mu$M for actins. Other parameters are taken from Table~\ref{table1}.} 
\label{excess_force}
\end{figure}

\begin{figure}
\centering
\includegraphics[scale = 0.65]{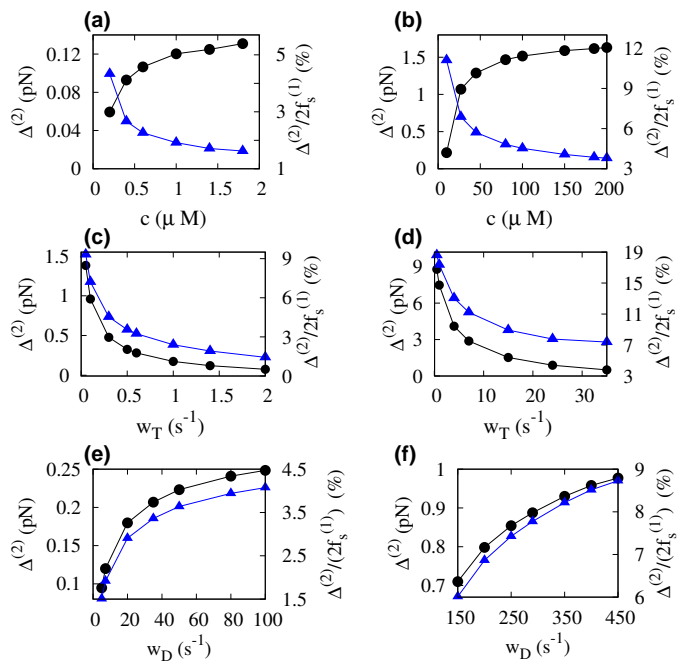}
\caption{ The deviation $\Delta^{(2)}$ (left $y$-label, $\bullet$) and percentage relative deviation $\Delta^{(2)}/(2 f_s^{(1)} )$ (right $y$-label, \tb{$\blacktriangle$}), for random hydrolysis, are plotted against: (a) concentration of G-actin monomers, (b) concentration of tubulin monomers,  (c) $w_T$ for actin,   (d) $w_T$ for microtubule, (e) $w_D$ for actin,   and (f) $w_D$ for microtubule. In figures (c)-(f) concentrations are $c=20\mu$M for microtubules and $c=1\mu$M for actins. All other parameters are as specified in Table~\ref{table1}} 
\label{fig5}
\end{figure}

We now investigate the dependence of the excess force on various parameters. For $N=2$, within the random model, we show the deviations 
($\Delta^{(2)}$) and percentage relative deviation ($\Delta^{(2)}/2 f_s^{(1)} $) as a function of free monomer concentration ($c$) for actin (Fig.~\ref{fig5}a) and microtubule (Fig.~\ref{fig5}b). The absolute deviation ($\Delta^{(2)}$) increases with $c$ and goes upto $0.13$ pN for actin and $1.61$ pN for microtubules. The percentage relative deviation is $\approx$ $5\%$ (for actin) and $12\%$ (for microtubules), at low $c$.
 Given that there is huge uncertainty in the estimate of $w_T$ for microtubule~\cite{Desai_Mitchison_MT:97}, and that in vivo proteins can regulate depolymerization rates, we systematically varied $w_T$.
 In Fig.~\ref{fig5}c-d we show that $\Delta^{(2)}$ increases rapidly with decreasing $w_T$ and can go up to $1.5$pN ($\Delta^{(2)}/2 f_s^{(1)} \approx 9\%$) for actin and $9$pN ($\Delta^{(2)}/2 f_s^{(1)} \approx 19\%$) for microtubules. 
{We did a similar study of $\Delta^{(2)}$ as a function of $w_D$  (see Fig. \ref{fig5}e-f), and find that   $\Delta^{(2)}$ increases with increasing $w_D$. The important thing to note is that $\Delta^{(2)}$ increases as we increase $w_D$ (at constant $w_T$) and decrease $w_T$ (at constant $w_D$), i.e. the magnitude of $\Delta^{(2)}$ is tied to the magnitude of difference of $w_D$ from $w_T$ (just as in our toy model in Sec. \ref{toy}).  This  also suggests that changes in depolymerization rates, typically regulated by  microtubule associated (actin binding) proteins in vivo, may  cause large variation in collective forces exerted by biofilaments.}

\begin{figure}
\centering
\includegraphics[scale = 0.5]{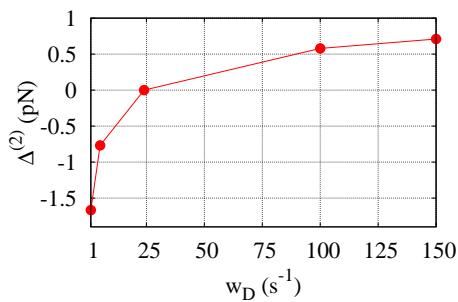}
\caption{ The deviation $\Delta^{(2)}$ for microtubule with random hydrolysis is plotted as a function of $w_D$ keeping fixed $w_T=24s^{-1}$, at a  concentration $c=20\mu$M (for other parameters see Table~\ref{table1}). Note that $\Delta^{(2)}=0$ at $w_D=w_T=24s^{-1}$.} 
\label{neg_delta2}
\end{figure}

{The case  $w_D=w_T$ effectively corresponds to the absence of switching, since dynamically  there is  no distinction between ATP/GTP-bound and ADP/GDP-bound subunits. When we move away from this point (i.e. when $w_D \neq w_T$), the effects of switching manifest. In these cases,  the hydrolysis process violates  detailed balance as it is  unidirectional: ATP/GTP $\rightarrow$ ADP/GDP conversion is never balanced by a reverse conversion ADP/GDP $\rightarrow$ ATP/GTP. Thus, similar to our toy model, we expect the non-equilibrium nature of switching (hydrolysis)  to be related to the phenomenon of excess force generation. In Fig. \ref{neg_delta2} we plot $\Delta^{(2)}$ as a function of $w_D$ (for smaller values of $w_D$ compared to Fig. \ref{fig5}f) for two-microtubule system, and find that indeed at the point $w_D=w_T$, $\Delta^{(2)}=0$. Moreover, for biologically impossible situations of $w_D < w_T$ (see Table~\ref{table1}), we find $\Delta^{(2)}<0$.}

\begin{figure}
\centering
\begin{subfigure}
\centering
\includegraphics[scale = 0.45]{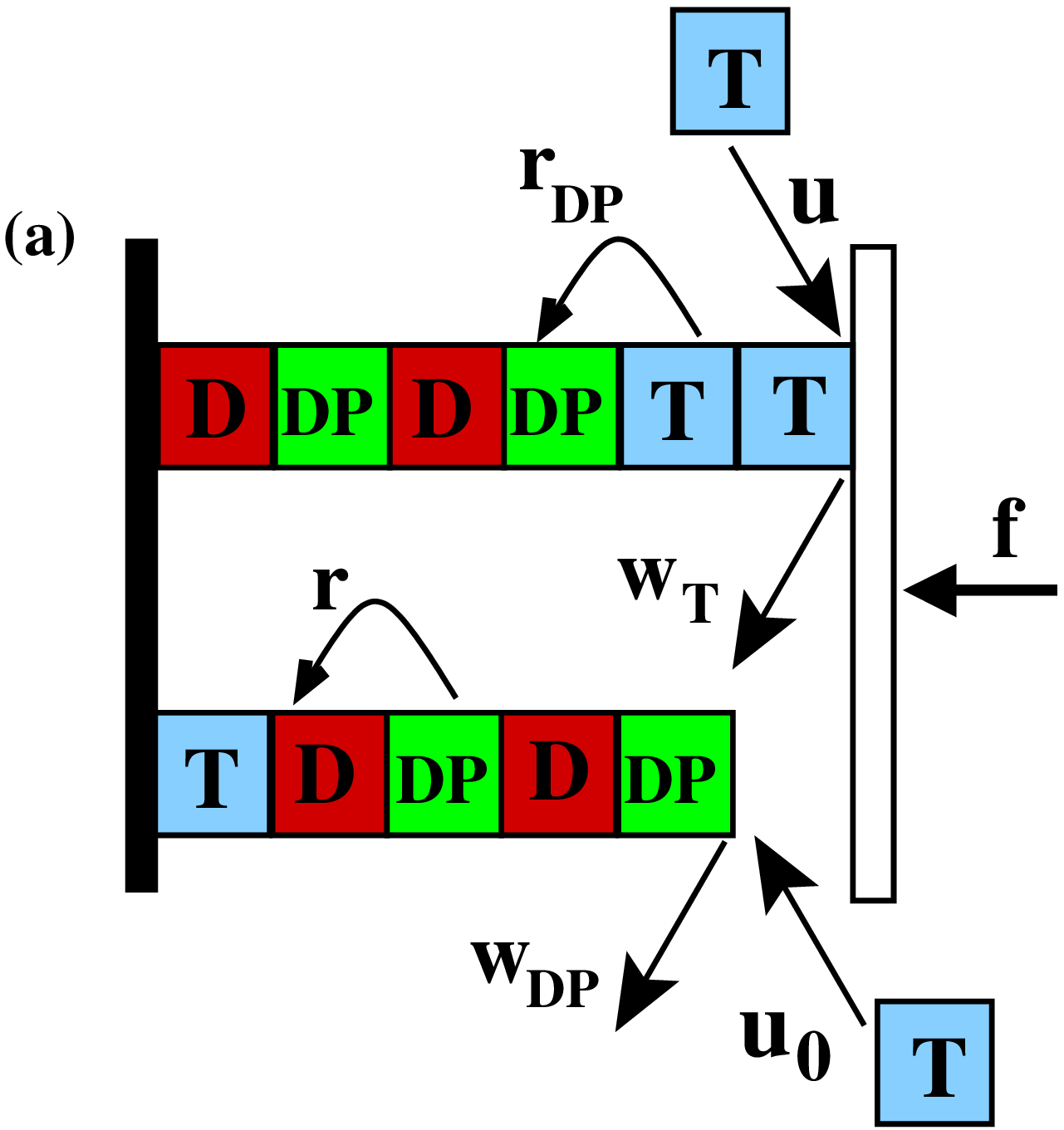}
\end{subfigure}
\begin{subfigure}
\centering
~~~~~\includegraphics[scale =0.57]{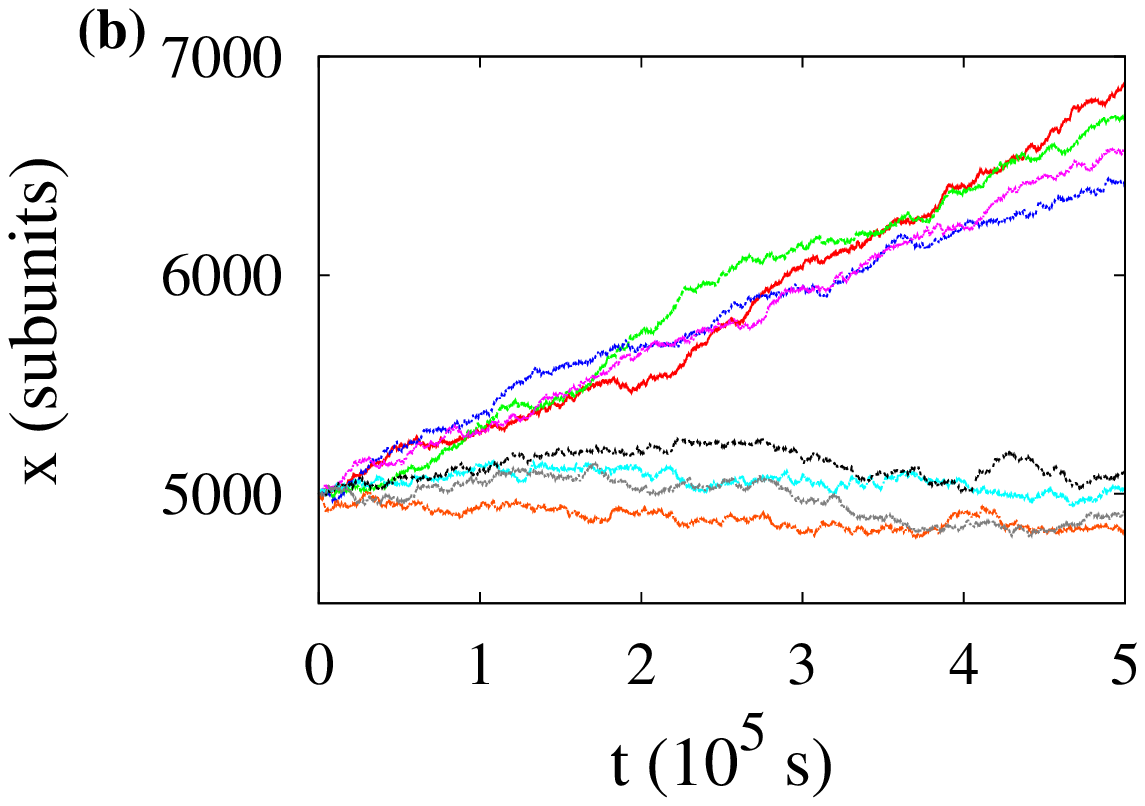}
\end{subfigure}
\caption{ (a) Schematic diagram of the three-state model with random hydrolysis, for  two actin filaments. Apart from ATP-bound (blue) and ADP-bound (red)  subunits, there is also an intermediate ADP-${\rm P_i}$-bound subunit (green). The corresponding rates are discussed in the text. (b)  Few traces of the wall position $x$ as a function of time for two actin filaments within the three state  model at a force $f=2 f_s^{(1)}$ (top), and  at   stall force $f_s^{(2)}$ (bottom).}
\label{3state}
\end{figure}

We now proceed to discuss the above phenomenon in further realistic variants of the random hydrolysis model. In reality actin hydrolysis involves two steps: ATP $\rightarrow$ ADP-${\rm P_i}$ $\rightarrow$ ADP \cite{Korn-Calier:87,vavylonis:05,Jegou2011}. Our two-state models above approximate this with the dominant rate limiting step of ${\rm P_i}$ release ~\cite{kolomeisky:06,Ranjith2009}. To test the robustness of our results, we study a more detailed ``three-state" model, which is defined by the following processes (as depicted in Fig. \ref{3state}a) and rates (taken from refs.~\cite{vavylonis:05,Jegou2011}): (i) addition of ATP-bound subunits  ($u_0=k_0 c = 11.6 s^{-1}$ with concentration $c=1 \mu$M), (ii) random ATP$\rightarrow$ADP-${\rm P_i}$ conversion ($r_{DP} = 0.3 s^{-1}$), (iii) random ADP-${\rm P_i}$$\rightarrow$ADP conversion ($r=0.003 s^{-1}$), (iv) dissociation of ATP-bound subunits ($w_T=1.4s^{-1}$), (v) dissociation of ADP-${\rm P_i}$-bound subunits ($w_{DP}=0.16s^{-1}$), and (vi) dissociation of ADP-bound subunits ($w_D=7.2s^{-1}$). For this model we first calculate the single-filament stall force $f_s^{(1)}=4.49$pN. Simulating this model for two actin filaments we find that the wall moves with a positive velocity at $f=2f_s^{(1)}$ (see Fig. \ref{3state}b-top) --- this proves $f_s^{(2)} > 2 f_s^{(1)}$. We then calculate the two-filament stall force $f_s^{(2)}=9.10$pN (also see Fig. \ref{3state}b-bottom for few traces of the wall position at stall). Consequently,  we obtain the excess force $\Delta^{(2)} = f_s^{(2)} - 2f_s^{(1)} = 0.12$pN --- this value is  same as that of the two-state random hydrolysis model (see Table \ref{compare}, actin).

In Ref \cite{Jegou2011}  a variant of the above three-state model is discussed where ADP-${\rm P_i}$$\rightarrow$ADP conversion happens at two different rates --- with a rate $r_{tip}$ at the tip, and with a rate $r$ in the bulk. In this model $r_{DP}=\infty$  i.e.  as soon as  an ATP subunit binds to a filament it converts to ADP-${\rm P_i}$ state --- this implies that effectively this model is a two-state model. We have simulated this model for actin filaments using the rates given in ref.~\cite{Jegou2011}: $u_0=k_0 c = 11.6 s^{-1}$  with concentration $c=1 \mu$M, $r_{tip}=1.8s^{-1}$, $r=0.007 s^{-1}$, $w_{DP}=0.16s^{-1}$,  and $w_D=5.8s^{-1}$. The stall forces for single filament and two filaments are $f_s^{(1)}=3.03$ pN and  $f_s^{(2)}=6.17$ pN respectively, implying $f_s^{(2)} > 2 f_s^{(1)}$. The resulting excess force $\Delta^{(2)} = 0.11$pN remains almost unchanged 
compared to the simple two-state random  hydrolysis model (see Table \ref{compare}, actin). 

{Although the multiple protofilament composition of actin and microtubules do not appear explicitly in any of the above models, we used effective subunit lengths to indirectly account for that. This drew from the fact that  
  the sequential hydrolysis  model can be  exactly mapped to a multi-protofilament model (called the ``one-layer" model) with strong inter-protofilament interactions ~\cite{kolomeisky:06, Ranjith2009} --- studies of two such composite filaments are discussed in \ref{1layer-seq}. We  further studied a new ``one-layer'' model with random hydrolysis in \ref{1layer-ran}, and  confirmed that simulation of the multi-protofilament model  yields similar results (even quantitatively) as the simpler random hydrolysis model discussed in this section.}

%
%

\section{Discussion and Conclusion}

 The study of force generation by cytoskeletal  filaments has been an active area of research for the last few years \cite{footer-dogterom:07,lacoste11, van-doorn:00, Laan-pnas:08,Ranjith2009,sanoop2013}. Earlier theories like our current work
have studied the phenomenon of  force generation by biofilaments and their dynamics in a theoretical picture  of  filaments  growing  against a constant applied force \cite{lacoste11, van-doorn:00,Ranjith2009}. These theories either neglected ATP/GTP hydrolysis, or looked at single ($N=1$) filament case, and hence outlined a 
 notion that  stall force of $N$ biofilaments is simply  $N$ times the stall force of a single one i.e. $f_s^{(N)}= N f_s^{(1)}$. In this paper, we theoretically show that this equality is {\it untrue} in the presence of hydrolysis. We find that $f_s^{(N)}> N f_s^{(1)}$ (for $N>1$), which is a manifestation of the non-equilibrium nature of the dynamics.  To establish this result beyond doubt, we showed it first analytically in a simple toy model which captures the essence of chemical switching. Then we proceeded to study the phenomenon in realistic random hydrolysis model and many of its variants (based on features of hydrolysis suggested from recent experimental literature). Even the sequential hydrolysis model is shown to exhibit similar phenomenon (see \ref{sequential}). Thus our extensive theoretical case studies suggest that the phenomenon of excess force generation is quite general and convincing.

 The question remains that how our results can be observed in suitably designed experiments in vitro. 
  There have been a few in vitro experiments which studied the phenomenon of collective force generation by biofilaments \cite{footer-dogterom:07, Laan-pnas:08}. In Ref. \cite{footer-dogterom:07} authors directly measured the force of parallel actin filaments by using an optical trap technique, and   found that the force generated by eight-actin filaments is much lesser than expected.   We would like to comment that  this experimental result can not be compared to  theoretical predictions like above due to the fact that the experiment is done under harmonic force, and not in a constant force ensemble as in theory. Secondly  the experimental filaments had buckling problems, which are unaccounted for in theory.
A later experiment \cite{Laan-pnas:08} on multiple microtubules (which were not allowed to buckle using a linear array of small traps) found that 
 the {\it most probable} values of forces generated by a bundle of microtubules appear as integral multiples of certain unit. This led to an interpretation that multiple microtubules generate force which grow linearly with filament number. We would like to comment that the theoretical stall forces mean {\it maximum} forces, which are not the most probable forces (as studied experimentally). Secondly like \cite{footer-dogterom:07}, experiment of \cite{Laan-pnas:08} also had harmonic forces, unlike constant forces in theories. 
To validate our claim of excess force generation,  new in vitro experiments should work within a constant force ensemble, ensure that  filaments do not buckle, and  the averaging of wall-velocity  is done over sufficiently long times, such that the effect of switching between heterogeneous states is truly sampled.

A simple way to prevent the buckling of the filaments is to keep their lengths short as the filaments under constant force will not bend below a critical length\cite{RobPhillips-book}. The estimates of the critical length for buckling, at stall force, for microtubule is $4-17\mu m$ (for $c=10-100\mu M$), and for actin is $0.5-3 \mu m$ (for $c=0.15 - 1 \mu M$). Since stall force does not depend on the length one can prevent buckling by choosing lengths well below the critical lengths of the filament, as done in~\cite{Laan-pnas:08}.
%
%
  
Apart from the direct measurement of the excess force $\Delta^{(N)}$, there is yet another way to check the validity of the relationship $f_s^{(N)}> N f_s^{(1)}$.
This relationship implies that an $N$-filament system would not stall at an applied force $f=N f_s^{(1)}$,  but would grow with positive velocity. For example, as seen in Fig. \ref {fig4}b, at $f=2 f_s^{(1)}$ two microtubules  with random hydrolysis have a velocity $\sim 1$ subunit/s --- equivalent to  a growth of $500$ nm in less than $15$ minutes.
In  Fig. \ref {fig4}c, one can see that, for two-actin within random hydrolysis, this velocity is $\sim 0.03$ subunits/s. This would imply a growth of $\sim 300$nm in $1$ hour. These velocities can  be even higher for other ATP/GTP concentrations. Such velocities (and the resulting length change) are considerable to be observed experimentally in a biologically relevant time scale --- thus the claim of the force equality violation may be validated. 

Even if the magnitude of the excess force is small, it can be crucial whenever there is a competition between two forces. For example, it is known that an active tensegrity picture~\cite{Ingbar:03} can explain cell shape, movement and many aspects of mechanical response. The core of the tensegrity picture is a force balance between growing microtubules and actin-dominated tensile elements Ð microtubule has to balance the compressive force exerted on them. So, in such a scenario, where two forces have to precisely balance, even a small change is enough to break the symmetry.

%

Our toy model has potential to go beyond this particular filament-growth problem: the model suggests that a simple two-state model with non-equilibrium switching can generate an interesting cooperative phenomenon and produce excess force/chemical potential than expected. In biology there are a number of non-equilibrium systems that switch between two (or multiple) states. For example, molecular motors, active channels across cell membrane etc. Following our finding, it will be of great interest to test whether such a cooperative phenomenon will arise in other biological and physical situations.

In this paper we did not address the issue of dynamic instability in the presence of force. This is an interesting problem in itself, and recent theoretical and experimental studies have addressed various aspects of this problem \cite{Laan-pnas:08, Kierfeld:2013}.
Our models are also capable of exhibiting this phenomenon, and interested reader may look at our recent work \cite{das-arxiv:2014} where we have studied
the collective catastrophes and cap dynamics
of multiple filaments under constant force, in detail, for the random
hydrolysis model, and compared our results to recent experiments
\cite{Laan-pnas:08}. The literature of two-state models \cite{Leibler:93,Kierfeld:2013} have shown that catastrophes arise in multiple filaments having force-dependent growth-to-shrinkage switching rates~\cite{Kierfeld:2013}. In microscopic models like ours, effective
force-dependent catastrophe rates emerge naturally; for example, in random hydrolysis model (see
\cite{Ranjith2012}), it is known that catastrophe rates are comparable to the results of Janson et al~\cite{jason-dogterom:03} and Drechsel et al~\cite{Drechsel:1992}. The toy model too can show dynamic instability with catastrophe and rescues (see \ref{appE}, and Fig.~\ref{appE:fig} (a)). Even with constant (force-independent) switching rates $k_{12}$ and $k_{21}$, the toy model
exhibits the phenomenon of force-dependent catastrophe; in Fig.~\ref{appE:fig} (b) we have shown that the rate ($k_{+-}$) of switching from growing state to shrinking state, computed from simulation, increases with force (also see \ref{appE}).
%
%
%
To test how the system behaves under explicit force-dependent switching, we made $k_{12} = k_{12}^{(0)} {\rm exp}({\tilde
f})$ with $w_2>w_1$;
%
we still find that $f_s^{(N)} >  N f_s^{(1)}$
(see \ref{appF} and Fig.  \ref{appF:fig}). Thus our
result of excess force generation is quite robust.
	One may also note that, in the random model $\Delta^{(N)}$ is roughly proportional to $N$ (Fig. \ref{excess_force}a and \ref{excess_force}b), while in the toy model the  $\Delta^{(N)}$
saturates (Fig. \ref{V-f} inset). This implies that, even though $f_s^{(N)} \neq N f_s^{(1)}$, we still have $f_s^{(N)}$ roughly scaling as $N$ for large $N$. 
The apparent similarity of this with the findings of Zelinski and Kierfeld~\cite{Kierfeld:2013}, where they show that the average polymerisation force of $N$ microtubules grows linearly with $N$ when  rescues are permitted, for filaments in harmonic trap, would be interesting to explore in detail in future.

In summary, we have studied collective dynamics of multiple biofilaments pushing against a wall and undergoing ATP/GTP hydrolysis. 
Quite contrary to the prevalent idea in the current
literature \cite{van-doorn:00,lacoste11,footer-dogterom:07}, we find that hydrolysis enhances the collective stall force compared to the sum of individual forces -- i.e. the equality $f_s^{(N)} = N f_s^{(1)}$ is untrue.
The understanding of the force equality was based on equilibrium arguments, 
and we have shown that  non-equilibrium processes of hydrolysis in bio-filaments lead to its violation. 

\ack 

We thank Mandar Inamdar for pointing out various non-equilibrium aspects of the system. We also thank J.-F. Joanny, A. Kolomeisky, D. Lacoste, S. Sen and T. Bameta for useful discussions, CSIR India (Dipjyoti Das, JRF award no.~09/087(0572)/2009-EMR-I) and IYBA,
Department of Biotechnology India (RP, No: BT/01/1YBAl2009) for financial support.

\appendix
\section*{Appendices}
\setcounter{section}{0}

\section{Toy model: single filament}
\label{toy-1filament}

The single-filament model discussed in the main text is shown schematically in Fig. \ref{1tube}. The state probabilities $P_1$ and $P_2$, defined in the main text, are related to $P_1(l,t)$ and $P_2(l,t)$, the probabilities of the filament of length $l$ being  in states $1$ and  $2$, respectively, at time $t$, as $P_1(t)=\sum_l P_1(l,t)$ and $P_2(t)=\sum_{l} P_2(l,t) $. The  probabilities obey

\bea
\frac{d P_1(l,t)}{ d t} &=& u P_1(l-1,t)+w_1 P_1(l+1,t)+k_{21}P_2(l,t) \nonumber\\&&- (u+w_1+k_{12})P_1(l,t) \label{Master1}\\
\frac{d P_2(l,t)}{ d t} &=& u P_2(l-1,t)+w_2 P_2(l+1,t)+k_{12}P_1(l,t)\nonumber\\
&& - (u+w_2+k_{21})P_2(l,t) \label{Master2}.\\ 
\nonumber
\eea

Using the above, after summing over all $l$, we get
\bea
\frac{d  P_1(t)}{d t} &=& k_{21}P_2(t)-k_{12}P_1(t) \label{switch1}\\
\frac{d  P_2(t)}{d t} &=& k_{12}P_1(t)-k_{21}P_2(t) \label{switch2}\\
\nonumber
\eea

\begin{figure}[ht]
\centering
\includegraphics[scale = 0.35]{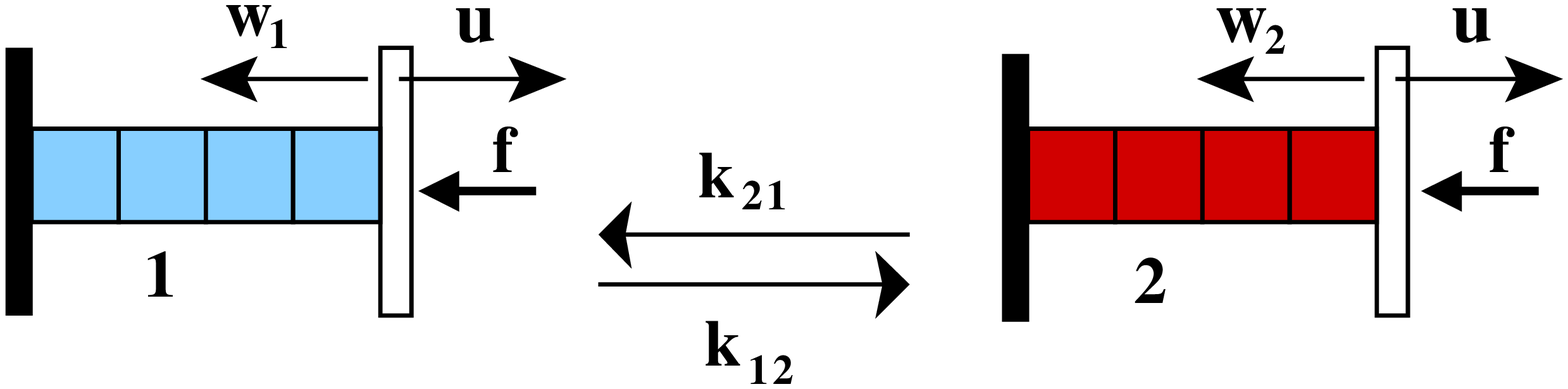}
\caption{Schematic diagram of single-filament model with switching between  states $1$ (blue) and $2$ (red). The polymerization rate is $u$, and the depolymerization rates are $w_1$ and $w_2$. The left wall is fixed, while the right wall is movable with a  force $f$ pressing against it.}
\label{1tube}
\end{figure}

The normalization condition is $\sum_{l=0}^{\infty}[ P_1(l,t)+P_2(l,t)] = P_1(t)+P_2(t)=1$. In steady state ($t \rightarrow \infty$) $P_1$ and $P_2$ become independent of time, and  both the  Eqs. (\ref{switch1},\ref{switch2}) give the same  detailed balance condition: 
\begin{eqnarray}
k_{12}P_1=k_{21}P_2
\label{bal}
\end{eqnarray}
Solving  Eq. \ref{bal} (along with the normalization condition) we  obtain: $P_1=k_{21}/(k_{12}+k_{21})$ and $P_2=k_{12}/(k_{12}+k_{21})$. The average position of the wall is given by $ \langle x(t) \rangle = \sum_{l=0}^{\infty} l [ P_1(l,t)+P_2(l,t)]$. So the velocity of the wall is:
\begin{eqnarray}
V^{(1)}&=& \frac{d \langle x(t) \rangle }{ d t} = \sum_{l=0}^{\infty} l [ \frac{d P_1(l,t)}{ d t}+\frac{d P_2(l,t)}{ d t}]\nonumber\\
&=& (u-w_1) \sum_{l} P_1(l,t) + (u-w_2) \sum_{l} P_2(l,t) \nonumber\\
&=& (u-w_1)  P_1(t) + (u-w_2)  P_2(t).\\
\nonumber
\end{eqnarray}
In the steady state ($t \rightarrow \infty$),
\begin{eqnarray}
V^{(1)}&=&[(u-w_1) k_{21} + (u - w_2) k_{12}]/(k_{12} + k_{21}),
\label{vel_1tube} 
\end{eqnarray}
which  is Eq. (1) in our main text.

\section{Toy model: Two filaments }
\label{toy-2filament}

\subsection{Derivation of the steady-state balance equations}
\label{toy-2filament-bal-eq}

In the two-filament model, we have four different states (see Fig. \ref{states} top). In the steady state, the system obeys probability flux balance conditions, which may be intuitively derived following  Fig.~\ref{states} (bottom). Below we provide a more systematic derivation of these equations starting from the microscopic  Master equations. Following the mathematical procedure of Ref. \cite{kolomeisky:05}, we define: $P_{ij}(l,l-k;t)$, the joint probability that, at time $t$, the top filament touching the wall (like in Fig.  \ref{2tube} (a)) is of length $l$  and in state $i$, and the bottom filament  of length $l-k$ ($l>k$) is in state $j$. Here $l$ and $k$ are natural numbers and $i,j$ =1 or 2.  We write (using the rates shown in Fig. \ref{2tube}) the following four Master equations satisfied by these probabilities corresponding to four different joint states (see Fig. \ref{states} top):

\bea
\frac{d  P_{11}(l,l-k;t)}{d t} &=& u P_{11}(l-1,l-k)+u_0 P_{11}(l,l-k-1)+w_{10}P_{11}(l,l-k+1) \nonumber\\ 
&&+w_1 P_{11}(l+1,l-k)+k_{21}(P_{12}(l,l-k)+P_{21}(l,l-k)) \nonumber \\ 
&& -(u+u_0+w_1+w_{10}+2k_{12})P_{11}(l,l-k) \label{M1} \\
\frac{d  P_{22}(l,l-k;t)}{d t} &=& u P_{22}(l-1,l-k)+u_0 P_{22}(l,l-k-1)+w_{20}P_{22}(l,l-k+1) \nonumber\\ 
&& +w_2 P_{22}(l+1,l-k)+k_{12}(P_{12}(l,l-k)+P_{21}(l,l-k))\nonumber\\
&&-(u+u_0+w_2+w_{20}+2k_{21})P_{22}(l,l-k)  \label{M2}\\
\frac{d  P_{12}(l,l-k;t)}{d t}&=& u P_{12}(l-1,l-k)+u_0 P_{12}(l,l-k-1)+w_{20}P_{12}(l,l-k+1) \nonumber\\
&&+w_1 P_{12}(l+1,l-k)+k_{12}P_{11}(l,l-k)+k_{21}P_{22}(l,l-k)\nonumber\\
&& -(u+u_0+w_1+w_{20}+k_{12}+k_{21})P_{12}(l,l-k)  \label{M3}\\
\frac{d  P_{21}(l,l-k;t)}{d t}&=& u P_{21}(l-1,l-k)+u_0 P_{21}(l,l-k-1)+w_{10}P_{21}(l,l-k+1) \nonumber\\
&& +w_2 P_{21}(l+1,l-k)+k_{12}P_{11}(l,l-k)+k_{21}P_{22}(l,l-k)\nonumber\\
&& -(u+u_0+w_2+w_{10}+k_{12}+k_{21})P_{21}(l,l-k) \label{M4}\\
\nonumber 
\eea

\begin{figure}
\centering
\includegraphics[scale = 0.2]{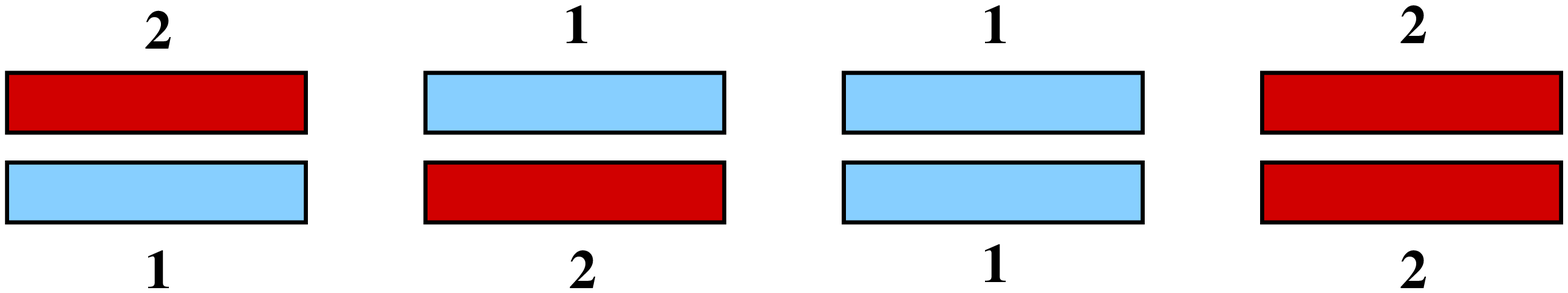}\\
\includegraphics[scale = 0.4]{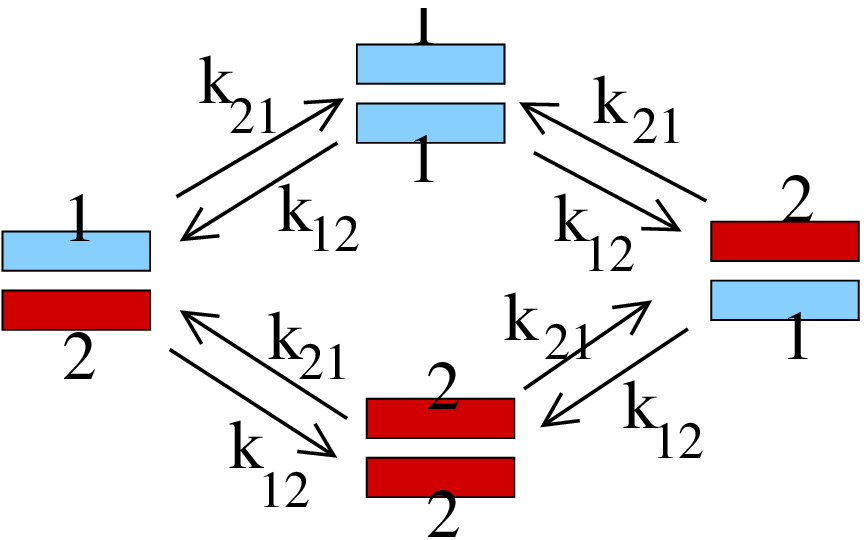}
\caption{ (top): Four possible joint states $\{2,1\}, \{1,2\},\{1,1\}$, and $\{2,2\}$ in the two-filament assembly are shown. Here joint state $\{i,j\}$ represents the situation where the top filament is in state $i$, and the bottom filament is in state $j$ ($i,j$ =1 or 2). (bottom): The schematic diagram indicating probability fluxes in and out of the joint states $\{i,j\}$.}
\label{states}
\end{figure}
\begin{figure}
\centering
\includegraphics[scale = 0.28]{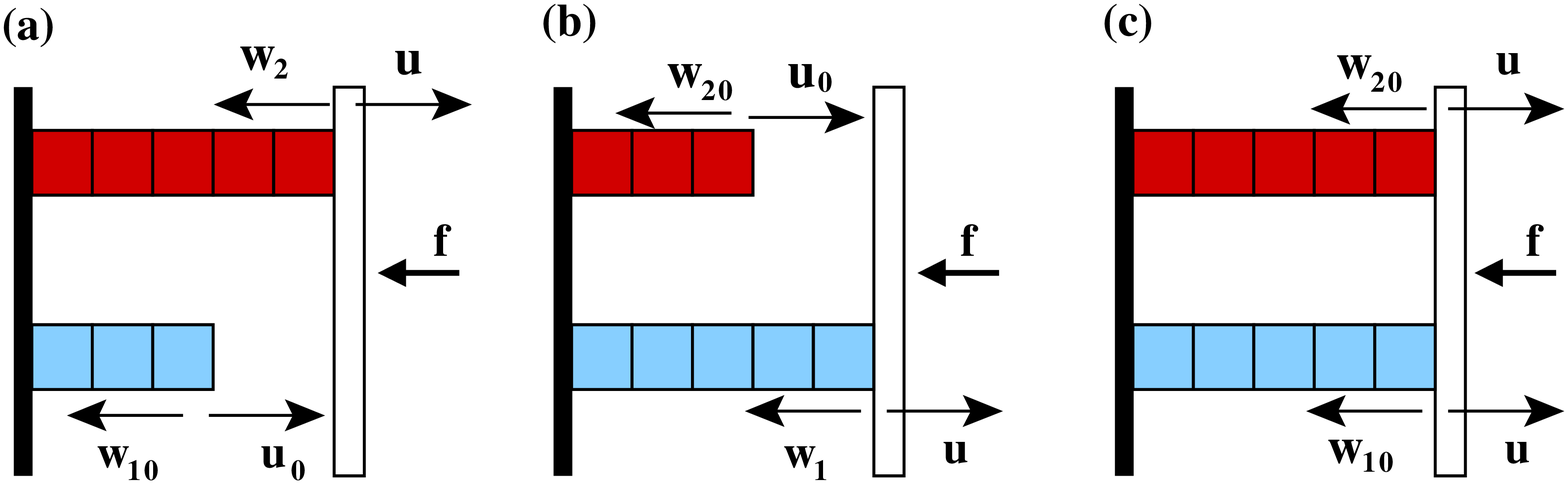}
\caption{ Schematic diagram of different length-configurations for two-filament model, when the filaments are in a joint state $\{2,1\}$. (a) The filament  in state $2$ is touching the wall. (b) The filament in state $1$ is touching the wall. (c) both the filaments have same length. Note that (in (c)) the depolymerization rates are force independent when both the filaments touch the wall.}
\label{2tube}
\end{figure}

Next, we define $P_{ij}(l-k,l+1;t)$, the  probability that, at time $t$, the top filament of length $l-k$ ($l\geq k$) is in state $i$, and the bottom filament of length $l+1$ touching the wall (see Fig.  \ref{2tube} (b)) is in state $j$. Here $l$ and $k$ are natural numbers and $i,j=$1 or 2.  These probabilities satisfy the following four Master equations:
\bea
\frac{d  P_{11}(l-k,l+1;t)}{d t} &=& u P_{11}(l-k,l)+w_{10} P_{11}(l-k+1,l+1)+u_{0}P_{11}(l-k-1,l+1)\nonumber\\
&&+w_1 P_{11}(l-k,l+2)+k_{21}(P_{12}(l-k,l+1)+P_{21}(l-k,l+1))\nonumber\\
&&-(u+u_0+w_1+w_{10}+2k_{12})P_{11}(l-k,l+1) \label{Mp1}\\ 
\frac{d  P_{22}(l-k,l+1;t)}{d t} &=& u P_{22}(l-k,l)+w_{20} P_{22}(l-k+1,l+1)+u_{0}P_{22}(l-k-1,l+1)\nonumber\\
&&+w_2 P_{22}(l-k,l+2)+k_{12}(P_{12}(l-k,l+1)+P_{21}(l-k,l+1))\nonumber\\
&&-(u+u_0+w_2+w_{20}+2k_{21})P_{22}(l-k,l+1)\label{Mp2} \\ 
\frac{d  P_{12}(l-k,l+1;t)}{d t} &=& u P_{12}(l-k,l)+w_{10} P_{12}(l-k+1,l+1)+u_{0}P_{12}(l-k-1,l+1)\nonumber\\
&&+w_2 P_{12}(l-k,l+2)+k_{12}P_{11}(l-k,l+1)+k_{21}P_{22}(l-k,l+1)\nonumber\\
&&-(u+u_0+w_2+w_{10}+k_{12}+k_{21})P_{12}(l-k,l+1) \label{Mp3}\\ 
\frac{d  P_{21}(l-k,l+1;t)}{d t} &=& u P_{21}(l-k,l)+w_{20} P_{21}(l-k+1,l+1)+u_{0}P_{21}(l-k-1,l+1)\nonumber\\
&&+w_1 P_{21}(l-k,l+2)+k_{12}P_{11}(l-k,l+1)+k_{21}P_{22}(l-k,l+1)\nonumber\\
&&-(u+u_0+w_1+w_{20}+k_{12}+k_{21})P_{21}(l-k,l+1)\label{Mp4} \\
\nonumber 
\eea
Next, let $P_{ij}(l,l;t)$ be the probability that, at time $t$, both filaments have same length $l$, and they are in a joint state $\{i,j\}$.  One such situation is shown in Fig.  \ref{2tube} (c). These probabilities satisfy the following Master equations:
\bea
\frac{d  P_{11}(l,l;t)}{d t} &=& u_0 P_{11}(l-1,l)+w_1 P_{11}(l,l+1)+u_0 P_{11}(l,l-1)+w_1 P_{11}(l+1,l) \nonumber\\ 
&&+k_{21}(P_{12}(l,l)+P_{21}(l,l))-2(u+w_{10}+k_{12})P_{11}(l,l) \label{M01} \\
\frac{d  P_{22}(l,l;t)}{d t} &=& u_0 P_{22}(l-1,l)+w_2 P_{22}(l,l+1)+u_0 P_{22}(l,l-1)+w_2 P_{22}(l+1,l) \nonumber\\
&& +k_{12}(P_{12}(l,l)+P_{21}(l,l))-2(u+w_{20}+k_{21})P_{22}(l,l) \label{M02}\\
\frac{d  P_{12}(l,l;t)}{d t}&=& u_0 P_{12}(l-1,l)+w_2 P_{12}(l,l+1)+u_0 P_{12}(l,l-1)+w_1 P_{12}(l+1,l) \nonumber\\
&& +k_{12} P_{11}(l,l)+ k_{21}P_{22}(l,l)-(2u+w_{10}+w_{20}+k_{12}+k_{21})P_{12}(l,l)\nonumber\\ 
\label{M03}\\
\frac{d  P_{21}(l,l;t)}{d t}&=& u_0 P_{21}(l-1,l)+w_1 P_{21}(l,l+1)+u_0 P_{21}(l,l-1)+w_2 P_{21}(l+1,l) \nonumber\\
&& +k_{12} P_{11}(l,l)+ k_{21}P_{22}(l,l)-(2u+w_{10}+w_{20}+k_{12}+k_{21})P_{21}(l,l) \nonumber\\
\label{M04}\\ \nonumber 
\eea
 We also define the probability of residency in joint state $\{i,j\}$ as: $P_{ij} \equiv \sum_{l,k} [P_{ij}(l,l-k; t)+P_{ij}(l-k,l+1;t)]$. The normalization of probabilities leads to $\sum_{{ \{i,j\}\\=1,2}} \sum_{l,k=0}^{\infty}[P_{ij}(l,l-k;t)+P_{ij}(l-k,l+1;t) ]= P_{11}+P_{22}+P_{12}+P_{21}=1$. Now one can take the sum over all $l$ and $k$ in the Master equations (\ref{M1} - \ref{M04}), and set the time-derivatives to zero to get the steady-state ($t\rightarrow \infty$) balance equations satisfied by the probabilities  $P_{ij}$:
\bea
 k_{21}(P_{12}+P_{21})&=& 2 k_{12}P_{11} \label{bal1},\\
 k_{12}(P_{12}+P_{21})&=& 2 k_{21}P_{22}\label{bal2},\\
 k_{12}P_{11}+k_{21}P_{22} &=& ( k_{12}+k_{21})P_{12},\label{bal3}\\
 k_{12}P_{11}+k_{21}P_{22} &=& ( k_{12}+k_{21})P_{21}.\label{bal4}\\
\nonumber
\eea
Note that one of the Eqs. (\ref{bal1}-\ref{bal4}) is redundant as it may be derived from the other three. From  Eqs. (\ref{bal3}) and (\ref{bal4}) we find the symmetric relationship $P_{12}=P_{21}$. 
Solving the above, along with the normalization condition, we get 
\bea
P_{11} &=& k_{21}^2/(k_{12} + k_{21})^2, \nonumber\\
 P_{22} &=& k_{12}^2/(k_{12} + k_{21})^2,\nonumber\\
P_{12} &=& P_{21} = k_{12}k_{21}/(k_{12} + k_{21})^2.
\label{pij-eq}
\eea

\subsection{Calculation of the velocity in the heterogeneous case }
\label{toy-2filament-v12}

The probabilities obtained above may be used to calculate the two-filament velocity $V^{(2)}=P_{11}v_{11}+P_{22}v_{22}+P_{12}v_{12}+P_{21}v_{21}$. The velocities $v_{11}$ and $v_{22}$ for homogeneous cases (i.e. when both filaments are  in the same state) can be directly read off from the result in Ref \cite{lacoste11} (see Eq. (2) in the main text). But we need to calculate afresh the velocities of heterogeneous cases, namely  $v_{12}$ and $v_{21}$, which are same by symmetry.

The velocity $v_{12}$ of the two-filament system with the top filament in state $1$ and the bottom in state $2$ is an average obtained by sampling all possible microscopic filament configurations in the $\{1,2\}$ state. In this state, let $p^{(i)}(k,t)$ be the probability that there is a gap of $k$ monomers between the two filaments and the filament which is in state $i$ ($i=$ 1 or 2), is touching the wall. Evidently we have: $p^{(1)}(k,t)=\sum_{l}  P_{12}(l,l-k;t)$ for $k\geq1$, and  $p^{(2)}(k,t)=\sum_{l}  P_{12}(l-k,l+1;t)$. We also define:  $p(0,t)=\sum_{l}  P_{12}(l,l;t)$, the probability of having zero gap between the filaments. Now since the filaments are not switching between the states (the top is always in $1$ and the bottom in $2$), Eqs. (\ref{M3}, \ref{Mp3}, \ref{M03}) for $P_{12}(l,l-k;t)$, $P_{12}(l-k,l+1;t)$, and $P_{12}(l,l;t)$ can be used after setting $k_{12}=k_{21}=0$. This leads to  the Master equations satisfied by $p^{(i)}(k,t)$ for $k\geq2$ :

\bea
\frac{d  p^{(1)}(k,t)}{d t} &=& (u_0+w_1)p^{(1)}(k+1)+(u+w_{20})p^{(1)}(k-1)-(u+u_0+w_{20}+w_1)p^{(1)} (k) \nonumber\\
\label{gap1}\\
\frac{d  p^{(2)}(k,t)}{d t} &=& (u_0+w_2)p^{(2)}(k+1)+(u+w_{10})p^{(2)}(k-1)-(u+u_0+w_{10}+w_2)p^{(2)}(k), \nonumber\\
\label{gap2}\\
\nonumber 
\eea
and for $k=1$ and $0$,
\bea
\frac{d  p^{(1)}(1,t)}{d t} &=& (u_0+w_1)p^{(1)}(2)+(u+w_{20})p(0)-(u+u_0+w_{20}+w_1)p^{(1)} (1) ~~~~~~~~~~~\label{gap3}\\
\frac{d  p^{(2)}(1,t)}{d t} &=& (u_0+w_2)p^{(2)}(2)+(u+w_{10})p(0)-(u+u_0+w_{10}+w_2)p^{(2)}(1), \label{gap4}\\
\frac{d  p(0,t)}{d t} &=& (u_0+w_1)p^{(1)}(1)+(u_0+w_{2})p^{(2)}(1)-(2u+w_{10}+w_{20})p (0). \label{gap5}\\
\nonumber 
\eea
In the steady state, the above Eqs. \ref{gap1}-\ref{gap5}, along with the normalization condition: $\sum_{n}p^{(1)}(k)+\sum_{n}p^{(2)}(k)+p(0)=1$, can be solved exactly and one gets the following  distributions for the gaps:
\begin{equation}
p^{(1)}(k)=p(0)\gamma_1^{k}~~,~~~p^{(2)}(k)=p(0)\gamma_2^{k},~~~~~~~~\rm{for}~~~ k\geq1.
\end{equation}
Here, $ p(0)=(1-\gamma_1)(1-\gamma_2)/(1-\gamma_1 \gamma_2)$,  $\gamma_1=(u+w_{20})/(u_0+w_1)$, and $\gamma_2=(u+w_{10})/(u_0+w_2)$. We must have $\gamma_1<1$ and $\gamma_2<1$ for the existence of the steady state. Now, the average velocity in the $\{1,2\}$ state is given by
\bea
v_{12}&=&(u-w_1) \sum_{k=1}^{\infty}p^{(1)}(k)+(u-w_2)\sum_{k=1}^{\infty}p^{(2)}(k)+2 u p(0)\nonumber\\
&=& p(0) \left[ \frac{\gamma_1(u-w_1)}{1-\gamma_1}+\frac{\gamma_2(u-w_2)}{1-\gamma_2}+2 u \right], 
\eea 
and  after simplification this leads to
\begin{equation}
v_{12}  = \frac{2u - (u+w_1)\gamma_1 - (u+w_2)\gamma_2 + (w_1+w_2)\gamma_1\gamma_2}{(1 - \gamma_1 \gamma_2)},
\label{v12_2tube}
\end{equation}
which is Eq. (3) in our main text.

\subsection{Two-filament stall force}
\label{toy-2filament-fs2}

Using the  explicit expressions for all  $P_{ij}$ (Eq.~\ref{pij-eq}) and $v_{11},v_{22}$ (Eq.~2, main text), $v_{12}$ (Eq.~\ref{v12_2tube}) we calculate the two-filament velocity: $V^{(2)}= P_{11}v_{11}+P_{22} v_{22}+2 P_{12} v_{12} = [k_{21}^2 v_{11} +    k_{12}^2 v_{22} + 2 k_{12} k_{21} v_{12}]/(k_{12}+k_{21})^2$.
At stall force $\tilde{f}=\tilde{f}_s^{(2)}$, $V^{(2)}$ is zero. So setting $V^{(2)}=0 $, and choosing $\delta=1$ (i.e. $u=u_0 {\rm e}^{-\tilde{f}}$, $w_1=w_{10},$ and $w_2=w_{20}$) we  obtain 
\begin{equation}
 \left. \frac{k_{21}^2 (u_0^2-w_{10}^2 {\rm e}^{\tilde{f}} )}{ u_0+(u_0+2 w_{10}){\rm e}^{\tilde{f}}} + \frac{k_{12}^2 (u_0^2-w_{20}^2 {\rm e}^{\tilde{f}})}{ u_0+(u_0+2 w_{20}){\rm e}^{\tilde{f}}} +  \frac{ 2 k_{12} k_{21} (u_0^2-w_{10} w_{20} {\rm e}^{\tilde{f}}) } {u_0+(u_0+w_{10}+w_{20}){\rm e}^{\tilde{f}} }\right |_{\tilde{f}=\tilde{f}_s^{(2)}} = 0,
\end{equation}
which is clearly a cubic equation in ${\rm e}^{\tilde{f}}$ whose only real root gives the  two-filament stall force $\tilde{f}_s^{(2)}$.



\section{Sequential hydrolysis model}
\label{sequential}

\begin{figure}
\centering
\includegraphics[scale = 0.5]{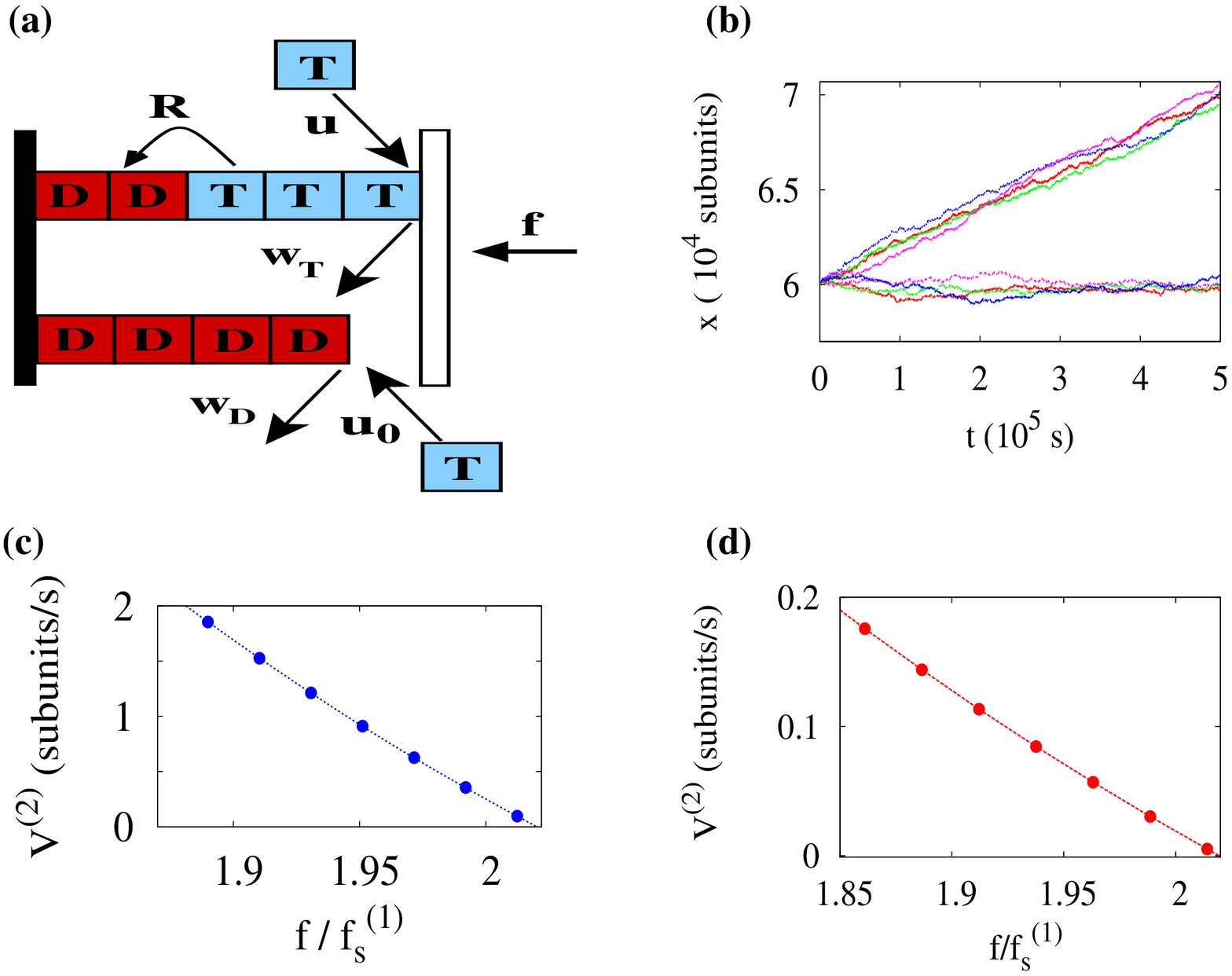}
\caption{(a) Schematic diagram of two-filament sequential hydrolysis model. ATP/GTP and ADP/GDP subunits are shown as letters `T' (blue) and `D' (red)  respectively. Here the switching ATP/GTP $\rightarrow$ ADP/GDP can happen only at the interface of ADP/GDP (bulk) and ATP/GTP (cap) region. For example, in the top filament the cap is made of three T subunits and the bulk is made of two D subunits. Various events (as described in the text below) are shown with arrows and corresponding rates. (b) Different traces of wall position ($x$) vs. time ($t$) of a two-microtubule system, for the sequential model at $f = 2f_s^{(1)}$ (top), and  at the stall force $f = f_s^{(2)}$ (bottom).  Scaled force-velocity relations of two-filament systems for (c) microtubules, and (d) actins. Parameters are specified in the text below. }
\label{seq}
\end{figure}

In this section we present the results for sequential hydrolysis model, while we focused on random hydrolysis model in the main text. The schematic diagram of a two-filament system with  sequential hydrolysis is shown in Fig. \ref{seq}(a). 
In this model,  polymerization of
filaments occurs with a rate $u = u_0 {\rm e}^{-\tilde{f}}$ (next to the wall) or
$u_0 $ (away from the wall). 
Note that the depolymerization rate is $w_T$ if there exists a finite ATP/GTP cap (like the top filament in Fig.~\ref{seq}(a)); otherwise it is  $w_D$ if the cap does not exist (like the bottom filament in Fig. \ref{seq}(a)). The hydrolysis rate (the rate of T becoming D) is $R$ and it can
happen only at the interface of the ADP/GDP(bulk)-ATP/GTP(cap)
regions. The switch ADP/GDP $\rightarrow$ ATP/GTP at the tip  can happen only by addition of free T monomers --- there is no direct conversion of ADP/GDP $\rightarrow$ ATP/GTP within a filament. For sequential hydrolysis the stall force of single filament is exactly known \cite{Ranjith2009}, which is
\begin{equation}
f_s^{(1)}= - \rm{ln} [(w_T+R)w_D / (w_D+R)u_0],
\label{fs1_seq}
\end{equation}
while for two filaments we need to calculate it numerically as no exact formula is  available.

Given the exactly known single filament stall force $f_s^{(1)}$ \cite{Ranjith2009}, we first apply a force $f = 2 f_s^{(1)}$ on a two-microtubule system and find   the wall moves with a positive velocity at $f = 2 f_s^{(1)}$ (see Fig.~\ref{seq}b (top)). The actual stall force  $f_s^{(2)}$ at which the wall halts on an average (Fig.\ref{seq}b (bottom)) is greater than $2 f_s^{(1)}$  --- this  can be clearly seen in the force-velocity plot for two microtubules, shown in Fig.\ref{seq}c. Similar, force-velocity curve for two actins is also shown in Fig.\ref{seq}d, where we again see that $f_s^{(2)}>2 f_s^{(1)}$.  We list in Table \ref{seq-table} the stall forces and excess forces at a fixed concentration  for actin and  microtubule within sequential hydrolysis. Actin parameters  are $c=1 \mu$M, $k_0=11.6 \mu$M$^{-1}s^{-1}$, $w_T=1.4 s^{-1}$, $w_D=7.2 s^{-1}$, and $R=0.3 s^{-1}$. For microtubule, these are: $c=100\mu$M, $k_0=3.2 \mu$M$^{-1}s^{-1}$, $w_T=24 s^{-1}$, $w_D=290 s^{-1}$, and $R=4 s^{-1}$.

\begin{table}
\caption{Stall forces and excess forces for sequential hydrolysis. Parameters are specified in the text below.}
\label{seq-table}
\begin{indented}
\item[]
\begin{tabular}{|c|c|c|c|}
\hline
& ~~~~$f_s^{(1)}$ (pN) & $f_s^{(2)}$  (pN) & $\Delta^{(2)}$ (pN) \\ \hline
Actin&~~~~$2.98$  & $6.02$ & $0.06$\\
\hline
Microtubule & ~~~~$16.74$ & $33.82$ & $0.34$\\ \hline
\end{tabular}
\end{indented}
\end{table}

\section{Multi-protofilament models}
\label{1layer}

\subsection{One-layer multi-protofilament  model with sequential hydrolysis}
\label{1layer-seq}

Microtubules and actin filaments are structures consisting of multiple proto-filaments that strongly interact with each other. Actin filaments are two-stranded helical polymers while microtubules are hollow cylinders made of 13 protofilaments~\cite{howard-book,Alberts-book, Desai_Mitchison_MT:97}. In this section we discuss the equivalence between the single-filament picture that we have been using, and the multiprotofilament nature of cytoskeletal filaments. In ref.~\cite{Ranjith2009}, it has been shown that, within the sequential hydrolysis, a multiprofilament model called ``one-layer" model can be exactly mapped to the single-filament picture we used. Below we discuss the one-layer model and show that our measured stall forces are  exactly the same as in the sequential hydrolysis model (see \ref{sequential} above).

One layer model makes use of two known experimental facts: (1) There is a strong lateral interaction between protofilaments (inter-protofilament interaction, which is as strong as $\approx -8 k_B$T for microtubules). (2) Each protofilament is shifted by a certain amount $\epsilon$ from its neighbor. We take  $\epsilon=b/m$ (see Fig. \ref{1_layer_seq}) where $b$ is the length of one tubulin/G-actin monomer, and $m$ is the number of protofilaments within one actin/microtubule ($m=2$ for actin, and $m=13$ for microtubule). Fact (1) would imply that any monomer binding on to a cytoskeletal filament (say, microtubule) would highly prefer a location that would form maximal lateral (inter-protifilament) bonds. This would lead to a situation where a growing cytoskeletal filament will be in a conformation where distance between any two protofilament tip will never be larger than $b$ (see Ref. \cite{kolomeisky:04,kolomeisky:05, kolomeisky:06} where this model is discussed in detail).

\begin{figure}
\centering
\includegraphics[scale = 0.4]{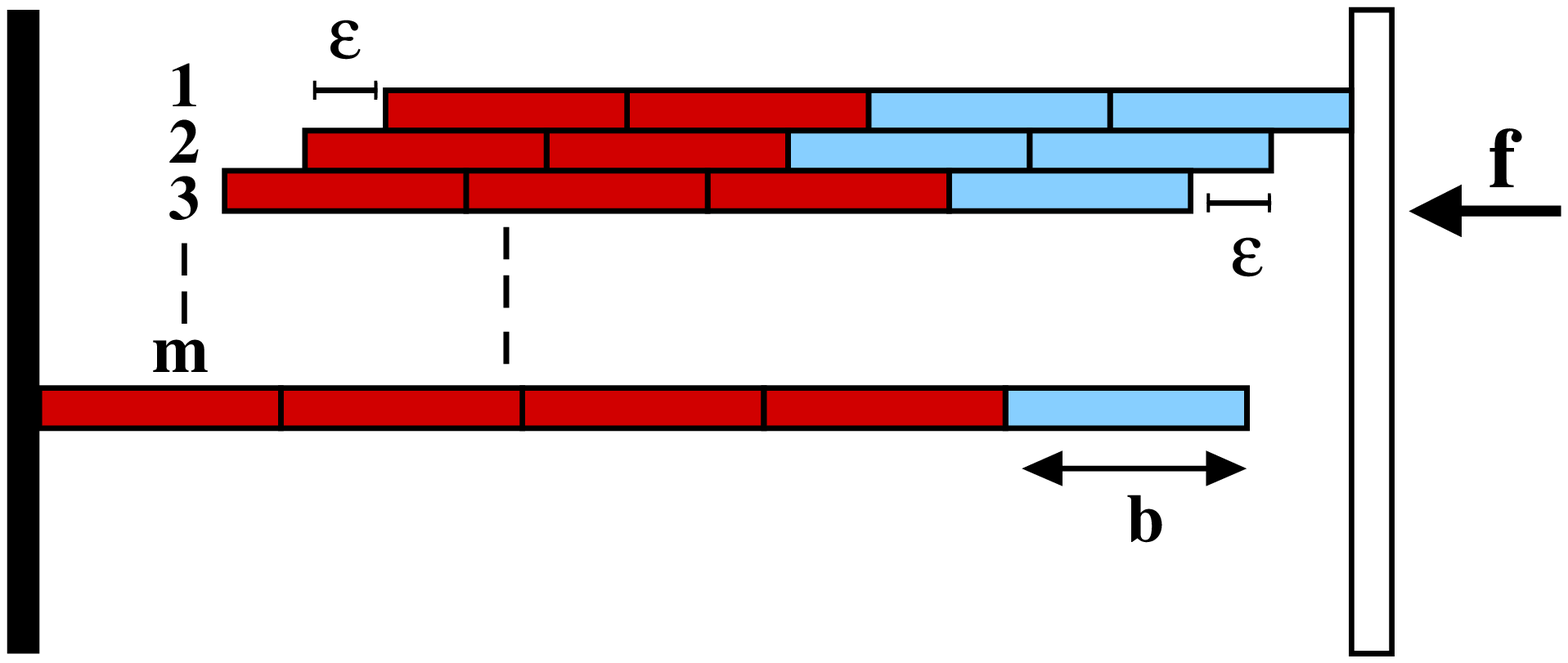}
\caption{Schematic diagram of a single filament made of $m$ protofilaments in one-layer model with sequential hydrolysis. Blue and red colors refer to ATP/GTP-bound and ADP/GDP-bound subunits respectively. Rules for the growth/shrinkage dynamics are discussed in the text. Dynamics happens only at one end (right) and the other end is inert. }
\label{1_layer_seq}
\end{figure}

The above-mentioned restrictions would lead to the following rules for growth dynamics:
(i) addition of a monomer can happen only at the most trailing tip at a rate $u=u_0 \exp(-{f}\epsilon/k_BT)$ -- for example, a monomer only can bind at protofilament $3$ in Fig. \ref{1_layer_seq}), (ii) dissociation of a monomer only takes place at most leading protofilament at a rate $w_T$ (when tip is ATP/GTP-bound) or $w_D$ (when tip is ADP/GDP-bound) -- for example, a monomer only can dissociate from protofilament $1$ in Fig. \ref{1_layer_seq}, and (iii) a hydrolysis event only happens at the most trailing T-D interface at a rate $R$ -- for example, a hydrolysis only takes place at protofilament $2$ in Fig. \ref{1_layer_seq}. It has been shown analytically \cite{Ranjith2009} that this one-layer model exactly maps for one actin filament ($m=2$) to a simple one-filament sequential hydrolysis model by taking $d=\epsilon=b/m$, where $d$ is the length of a subunit in sequential hydrolysis model. Note that this mapping is expected since in the one-layer model the right wall (see Fig. \ref{1_layer_seq}) only moves by an amount of  $\epsilon$ after each association/dissociation event. We further numerically find that this mapping exactly works for  one microtubule and two microtubules and actin filaments. All the stall forces $f_s^{(1)}$,  $f_s^{(2)}$, and thus the excess force $\Delta^{(2)}$ are exactly same in real units (with the above mapping) as in sequential hydrolysis model (see Table \ref{seq-table}).

\subsection{One-layer multi-protofilament  model with random hydrolysis}
\label{1layer-ran}

\begin{figure}
\centering
\includegraphics[scale = 0.4]{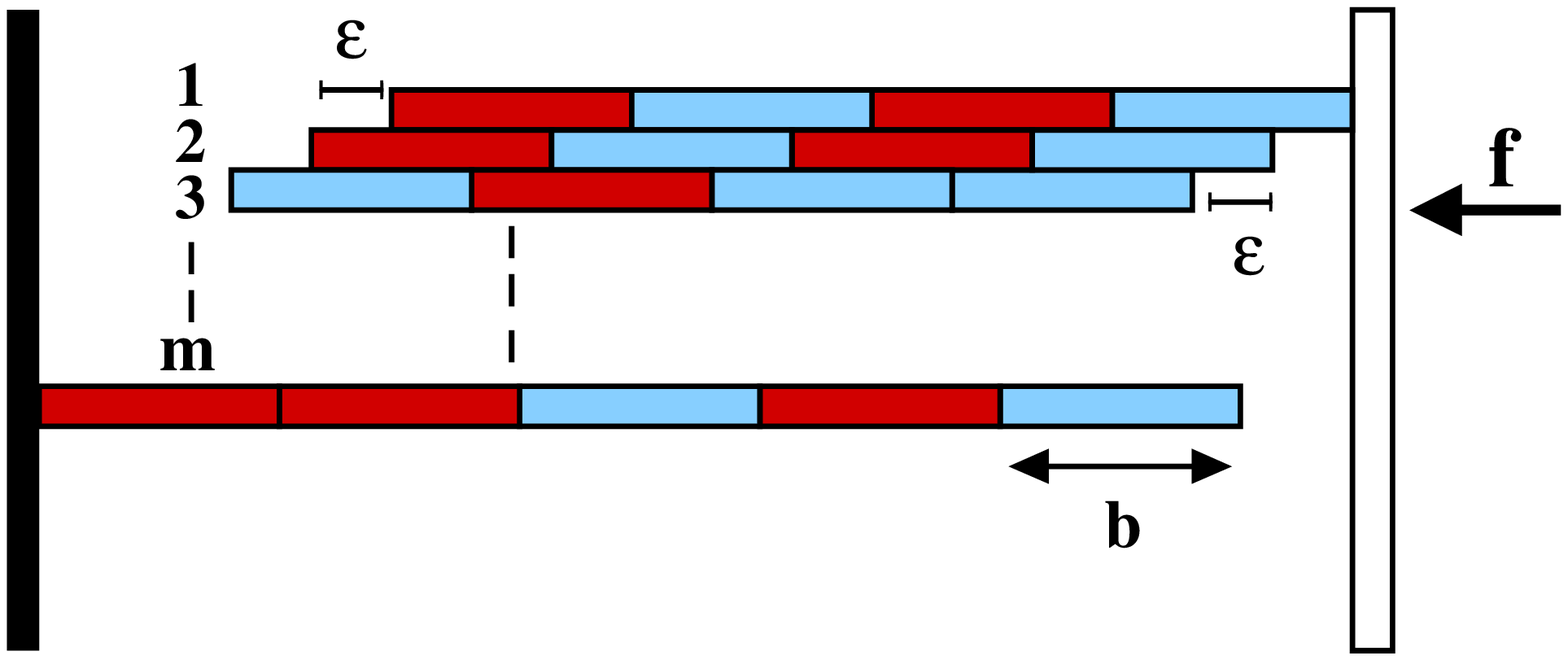}
\caption{Schematic diagram of a single filament made of $m$ protofilaments in one-layer model with random hydrolysis. Blue and red colors refer to ATP/GTP-bound and ADP/GDP-bound subunits respectively. Rules for the growth/shrinkage dynamics are discussed in the text below.}
\label{1_layer_rand}
\end{figure}

\begin{table}
\caption{Comparison of stall force and excess force values between simplified random hydrolysis model (Fig.~\ref{random-hydro}) and  One-layer (multi-protofilament)  random hydrolysis  model (Fig. \ref{1_layer_rand}), at fixed concentrations $c=100 \mu$M for microtubules and  $c=1 \mu$M for actins. Other parameters are specified in Table \ref{table1}.}
\label{compare}
\begin{indented}
\item[]
\centering
\begin{tabular}{|l|l|l|l||l|l|l|} \hline
~ & \multicolumn{3}{c||} {Simplified random hydrolysis} & \multicolumn{3}{c|}{One-layer (multi-protofilament)} \\ \cline{2-7}
~ & ~~~~$f_s^{(1)}$ (pN) & $f_s^{(2)}$  (pN) & $\Delta^{(2)}$ (pN) & ~~~~$f_s^{(1)}$  (pN)& $f_s^{(2)}$ (pN) & $\Delta^{(2)}$ (pN)\\ \hline
Actin& ~~~~$3.13$  & $6.38$ & $0.12$  & ~~~~$3.14$ & $6.38$ & $0.10$ \\ \hline
Microtubule & ~~~~$16.75$ & $35.01$  & $1.51$ & ~~~$17.06$ & $35.06$ & $0.94$ \\ \hline
\end{tabular}
\end{indented}
\end{table}

In the spirit of ``one-layer'' sequential hydrolysis model  \cite{kolomeisky:04,kolomeisky:05, kolomeisky:06} we propose a one-layer version within the random hydrolysis. We consider a biofilament made of $m$ protofilaments, and  each protofilament is shifted by an amount $\epsilon=b/m$ from its neighbour (see Fig. \ref{1_layer_rand}). Here $b$ is the length of a Tubulin/G-actin monomer. To simulate  the system we apply the following rules for polymerization and depolymerization and hydrolysis dynamics: (i) addition of a monomer can happen only at the most trailing protofilament tip with a rate $u=u_0 \exp(-f\epsilon/k_BT)$ -- for example, a monomer can bind only at protofilament $3$ in Fig. \ref{1_layer_rand}, (ii) dissociation of a monomer takes place only at the most leading protofilament with a rate $w_T$ (when tip is ATP/GTP-bound) or $w_D$ (when tip is ADP/GDP-bound) -- for example, a monomer can dissociate only from protofilament $1$ in Fig. \ref{1_layer_rand}, and (iii) a random hydrolysis event happens only at that protofilament which has maximum number of ATP/GTP-bound subunits with a rate $r$ -- for example, a hydrolysis event takes place  only at protofilament $3$ in Fig. \ref{1_layer_rand} (at any random location). Numerically we find that the results for stall forces and excess forces in this model are very close to that of the random hydrolysis model (see Table \ref{compare}) -- to make the correspondence, we set the subunit-length $d=\epsilon=b/m$ in the random hydrolysis model. Thus, $d=8$nm$/13=0.6$nm for microtubule and  $d=5.4$nm$/2=2.7$nm for actin within random hydrolysis model.

\section{ Force-dependence of the growth-to-shrinkage switching rate  within the toy model}
\label{appE}

\begin{figure}
\centering
\includegraphics[scale = 0.45]{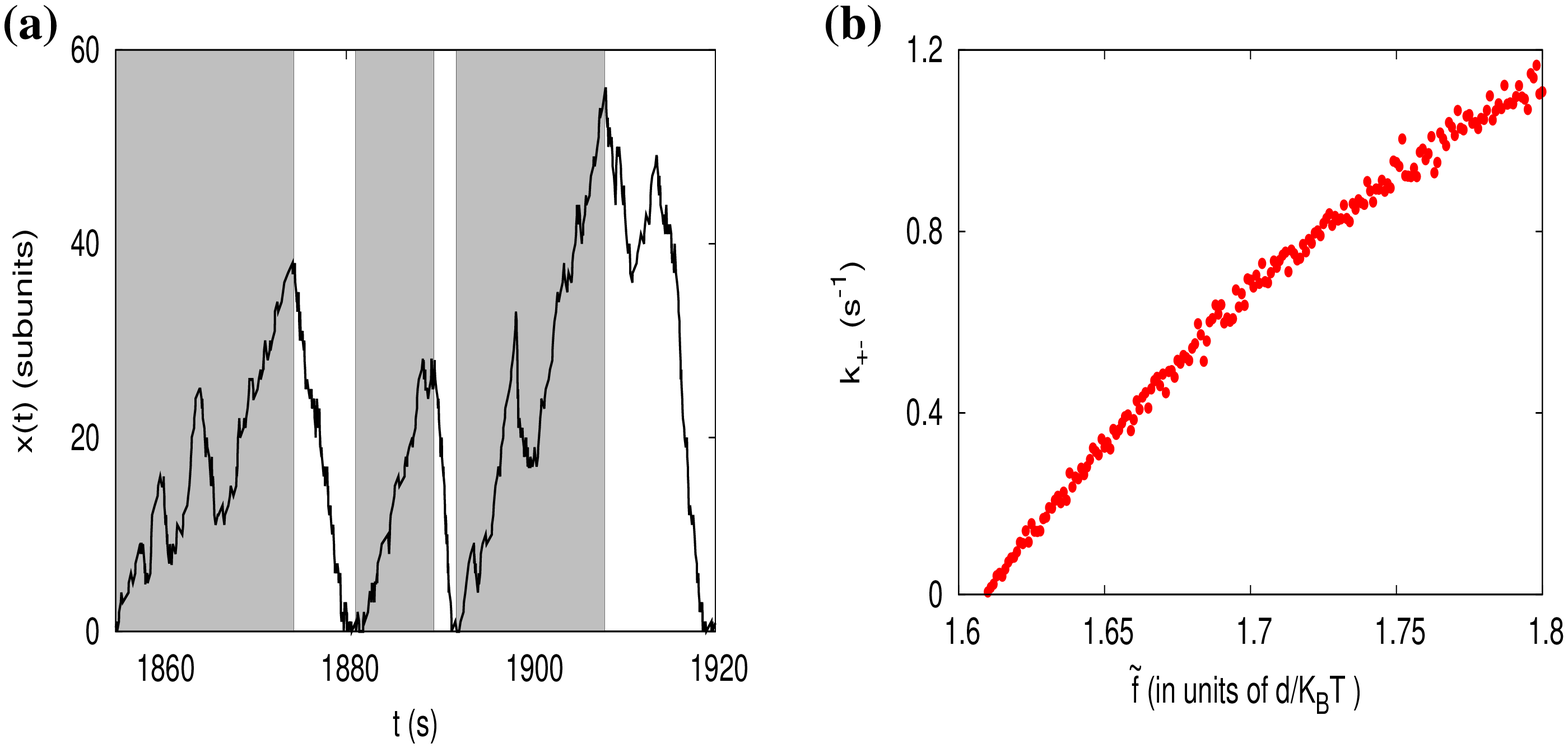}
\caption{ (a) Time-trace of the wall position showing  catastrophes within the toy model at a (dimensionless) force $\tilde{f}=1.85 >\tilde{f}_s^{(1)}$ ($\tilde{f}_s^{(1)}=1.61$ in this case). The regions shaded grey correspond to the growing states and is used to compute $T_{+}$. (b) The ``growth-to-shrinkage'' switching rate $k_{+-} (=1/\langle T_{+} \rangle)$ versus scaled force $\tilde{f}$ in the toy model. For both figures, the parameters are: $u_0=40 s^{-1}, w_{10}=1 s^{-1}, w_{20}=15 s^{-1}, k_{12}=k_{21}=0.5 s^{-1}$, and $\delta=1$.}
\label{appE:fig}
\end{figure}

As discussed in the literature~\cite{Leibler:93}, microtubule dynamics can be classified into two dynamical phases -- (i) bounded growth phase (average filament velocity $v=0$) and (ii) unbounded growth phase ($v>0$). For forces greater than the stall force  ${f}_s^{(1)}$, the  filament length  fluctuates around a constant -- this is the bounded phase. While, at the unbounded growth phase the  average filament length increases with time. To check whether our toy model shows catastrophes, we simulated the dynamics of a filament in the bounded growth phase. A typical time trace of the wall position ($x$ vs $t$) is shown in Fig.~\ref{appE:fig}a -- we clearly see the filament collapsing to zero length frequently. 
Following Fig.~\ref{appE:fig}a, we define a ``peak'' to be the highest value of  $x$ between two successive zero values. We then define the growth time ($T_{+}$) as the time it takes to reach a ``peak'' starting from the preceding zero (see the regions shaded grey in Fig. \ref{appE:fig}a).  We construct a  switching rate from growth to shrinkage  as $k_{+-}=1/\langle T_{+} \rangle$, where $\langle T_{+} \rangle$ is measured by averaging over a long time window. In Fig. \ref{appE:fig}b we show that the rate $k_{+-}$ increases with the force.

\section{A variant of the toy model with force-dependent switching rate between  depolymerization states}
\label{appF}

\begin{figure}
\centering
\includegraphics[scale = 0.3,angle=-90]{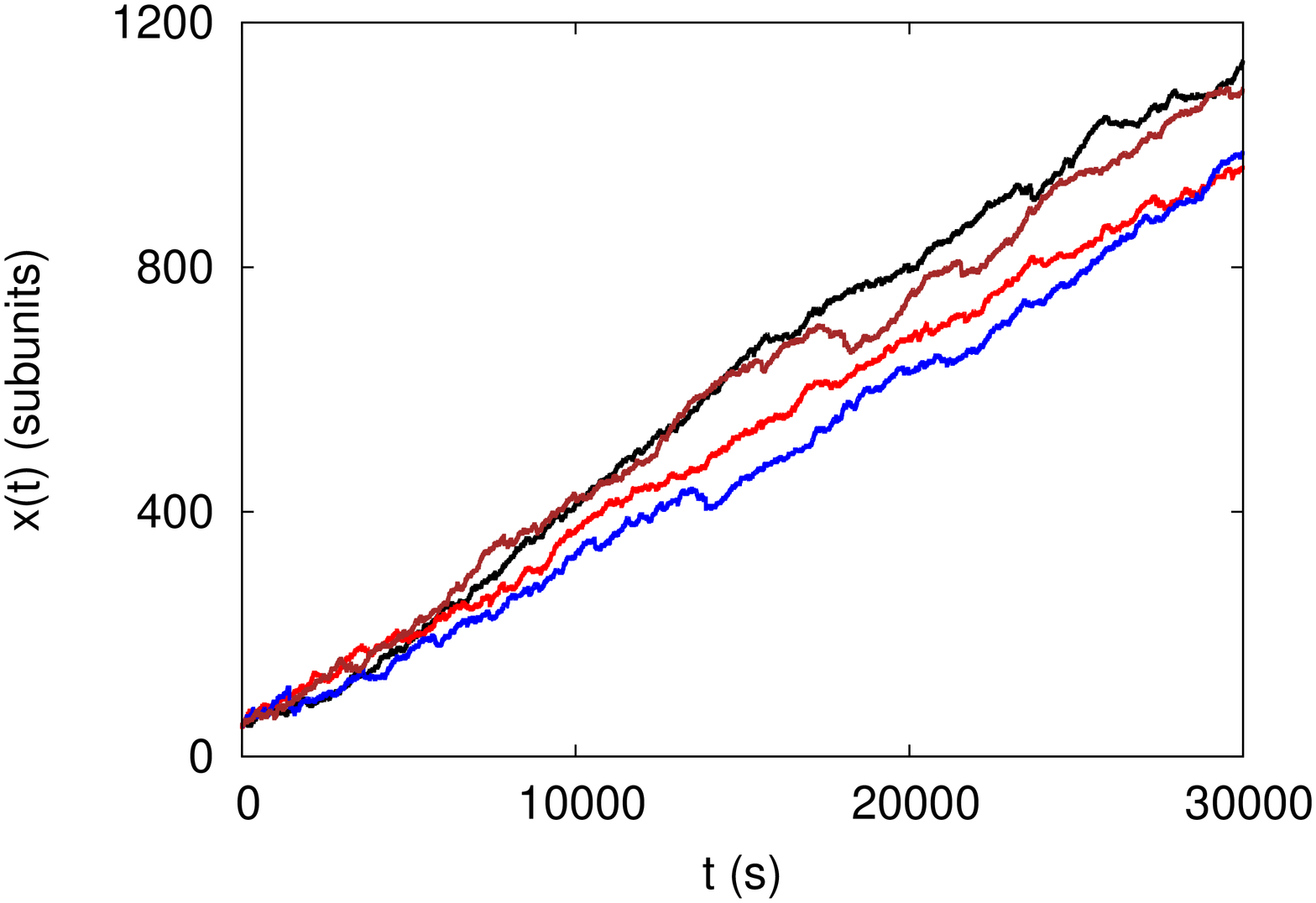}
\caption{Different traces of the wall position (x) vs. time (t) of a two filament system at a dimensionless force $\tilde{f}=2\tilde{f}_s^{(1)}$, for a variant of the toy model having $k_{12} = k_{12}^{(0)} {\rm exp}(\tilde{f})$. The parameters are: $u_0=40 s^{-1}, w_{10}=1 s^{-1}, w_{20}=15 s^{-1}, k_{12}^{(0)}=0.5 s^{-1}, k_{21}=0.5 s^{-1}$, and $\delta=1$.}
\label{appF:fig}
\end{figure}

The rates $k_{12}$ and  $k_{21}$, in the toy model we studied, represent the switching between two distinct ``depolymerization states''. As a result, in these states a filament has different growth velocities. We now make  the switching rate from low-depolymerisation-rate state (high velocity state) to high-depolymerisation-rate state (low velocity state) force dependent, namely $k_{12} = k_{12}^{(0)} {\rm exp}(\tilde{f})$. With this modification the exact value of one filament stall force can be calculated from Eq. \ref{V1} (main text) as  $\tilde{f}_s^{(1)} = 3.277$ for the parameters specified in the caption of Fig.~\ref{appF:fig}. We then apply a force $\tilde{f}=2\tilde{f}_s^{(1)}$ on a two filament system, and show in Fig. \ref{appF:fig} that the wall moves forward with a positive velocity, implying $\tilde{f}_s^{(2)} > 2\tilde{f}_s^{(1)}$. The corresponding excess force $\tilde{\Delta}^{(2)}= \tilde{f}_s^{(2)} - 2\tilde{f}_s^{(1)} = 0.4$.

\section*{References}
\bibliographystyle{unsrt}

\begin{thebibliography}{50}

\bibitem{howard-book}
J.~Howard.
\newblock {\em Mechanics of Motor Proteins and the Cytoskeleton}.
\newblock Sinauer Associates, Inc., Massachusetts, 2001.

\bibitem{Alberts-book}
Bruce Alberts, Alexander Johnson, Julian Lewis, Martin Raff, Keith Roberts, and
  Peter Walter.
\newblock {\em {Molecular Biology of the Cell}}.
\newblock Garland Science, New York, 4 edition, 2002.

\bibitem{Pollard1986}
Thomas~D Pollard.
\newblock {Rate constants for the reactions of ATP- and ADP-actin with the ends
  of actin filaments.}
\newblock {\em J. Cell Bio.}, 103:2747--2754, 1986.

\bibitem{Desai_Mitchison_MT:97}
A.~Desai and T.~J. Mitchison.
\newblock Microtubule polymerization dynamics.
\newblock {\em Annu. Rev. Cell. Dev. Biol.}, 13(1):83--117, 1997.

\bibitem{Korn-Calier:87}
E~D Korn, M~F Carlier, and D~Pantaloni.
\newblock {Actin polymerization and ATP hydrolysis}.
\newblock {\em Science}, 238:638--644, 1987.

\bibitem{vavylonis:05}
Dimitrios Vavylonis, Qingbo Yang, and Ben O'Shaughnessy.
\newblock {Actin polymerization kinetics, cap structure, and fluctuations}.
\newblock {\em Proc. Natl. Acad. Sci. USA}, 102(24):8543--8548, 2005.

\bibitem{Jegou2011}
Antoine J\'{e}gou, Thomas Niedermayer, J\'{o}zsef Orb\'{a}n, Dominique Didry,
  Reinhard Lipowsky, Marie-France Carlier, and Guillaume Romet-Lemonne.
\newblock {Individual Actin Filaments in a Microfluidic Flow Reveal the
  Mechanism of ATP Hydrolysis and Give Insight Into the Properties of
  Profilin}.
\newblock {\em PLoS Biology}, 9(9):e1001161, September 2011.

\bibitem{footer-dogterom:07}
Matthew~J. Footer, Jacob W.~J. Kerssemakers, Julie~A. Theriot, and Marileen
  Dogterom.
\newblock Direct measurement of force generation by actin filament
  polymerization using an optical trap.
\newblock {\em Proc. Natl. Acad. Sci. USA}, 104(7):2181--2186, 2007.

\bibitem{lacoste11}
Kostas Tsekouras, David Lacoste, Kirone Mallick, and Jean-François Joanny.
\newblock Condensation of actin filaments pushing against a barrier.
\newblock {\em New J. Phys.}, 13(10):103032, 2011.

\bibitem{hill:81}
T~L Hill.
\newblock {Microfilament or microtubule assembly or disassembly against a
  force}.
\newblock {\em Proc. Natl. Acad. Sci. USA}, 78(9):5613--5617, 1981.

\bibitem{kolomeisky:06}
Evgeny~B. Stukalin and Anatoly~B. Kolomeisky.
\newblock {ATP Hydrolysis Stimulates Large Length Fluctuations in Single Actin
  Filaments}.
\newblock {\em Biophys. J.}, 90(8):2673--2685, 2006.

\bibitem{hill:85}
D~Pantaloni, T~L Hill, M~F Carlier, and E~D Korn.
\newblock {A model for actin polymerization and the kinetic effects of ATP
  hydrolysis}.
\newblock {\em Proc. Natl. Acad. Sci. USA}, 82(21):7207--7211, 1985.

\bibitem{Leibler-cap:96}
Henrik Flyvbjerg, Timothy~E. Holy, and Stanislas Leibler.
\newblock Microtubule dynamics: Caps, catastrophes, and coupled hydrolysis.
\newblock {\em Phys. Rev. E}, 54(5):5538--5560, Nov 1996.

\bibitem{Ranjith2009}
P.~Ranjith, D.~Lacoste, K.~Mallick, and J-F. Joanny.
\newblock {Nonequilibrium Self-Assembly of a Filament Coupled to ATP/GTP
  Hydrolysis}.
\newblock {\em Biophys. J.}, 96:2146--2159, 2009.

\bibitem{Ranjith2012}
Ranjith Padinhateeri, Anatoly~B Kolomeisky, and David Lacoste.
\newblock {Random Hydrolysis Controls the Dynamic Instability of Microtubules}.
\newblock {\em Biophys. J.}, 102:1274--1283, 2012.

\bibitem{Manoj:2013}
V.~Jemseena and Manoj Gopalakrishnan.
\newblock Microtubule catastrophe from protofilament dynamics.
\newblock {\em Phys. Rev. E}, 88:032717, Sep 2013.

\bibitem{kolomeisky:05}
E.~B. Stukalin and A.~B. Kolomeisky.
\newblock Polymerization dynamics of double-stranded biopolymers: chemical
  kinetic approach.
\newblock {\em J. Chem. Phys.}, 122(10):104903, 2005.

\bibitem{van-doorn:00}
G.~Sander van Doorn, Catalin Tanase, Bela~M. Mulder, and Marileen Dogterom.
\newblock On the stall force for growing microtubules.
\newblock {\em Eur. Biophys. J.}, 20:2--6, 2000.

\bibitem{Kierfeld:2011}
J.~Krawczyk and Jan Kierfeld.
\newblock Stall force of polymerizing microtubules and filament bundles.
\newblock {\em Euro. Phys. Lett.}, 93:28006, 2011.

\bibitem{Laan-pnas:08}
Liedewij Laan, Julien Husson, E.~Laura Munteanu, Jacob W.~J. Kerssemakers, and
  Marileen Dogterom.
\newblock Force-generation and dynamic instability of microtubule bundles.
\newblock {\em Proc. Natl. Acad. Sci. USA}, 105(26):8920--8925, 2008.

\bibitem{Kierfeld:2013}
Bj\"orn Zelinski and Jan Kierfeld.
\newblock Cooperative dynamics of microtubule ensembles: Polymerization forces
  and rescue-induced oscillations.
\newblock {\em Phys. Rev. E}, 87:012703, 2013.

\bibitem{mitchison:2009}
Hao~Yuan Kueh and Timothy~J. Mitchison.
\newblock {Structural Plasticity in Actin and Tubulin Polymer Dynamics}.
\newblock {\em Science}, 325:960--963, 2009.

\bibitem{Dogterom-yurke:97}
Marileen Dogterom and Bernard Yurke.
\newblock {Measurement of the Force-Velocity Relation for Growing
  Microtubules}.
\newblock {\em Science}, 278(5339):856--860, 1997.

\bibitem{sumedha}
Sumedha, Michael~F. Hagan, and Bulbul Chakraborty.
\newblock Prolonging assembly through dissociation: A self-assembly paradigm in
  microtubules.
\newblock {\em Phys. Rev. E}, 83:051904, 2011.

\bibitem{Antal-etal-PRE:07}
T.~Antal, P.~L. Krapivsky, S.~Redner, M.~Mailman, and B.~Chakraborty.
\newblock Dynamics of an idealized model of microtubule growth and catastrophe.
\newblock {\em Phys. Rev. E.}, 76(4):041907, 2007.

\bibitem{Li:2009}
Xin Li, Jan Kierfeld, and Reinhard Lipowsky.
\newblock Actin polymerization and depolymerization coupled to cooperative
  hydrolysis.
\newblock {\em Phys. Rev. Lett.}, 103:048102, Jul 2009.

\bibitem{Kolomeisky:2013}
Xin Li and Anatoly~B. Kolomeisky.
\newblock Theoretical analysis of microtubules dynamics using a
  physical-chemical description of hydrolysis.
\newblock {\em The Journal of Physical Chemistry B}, 117(31):9217--9223, 2013.

\bibitem{Ranjith2010}
P.~Ranjith, K.~Mallick, J-F. Joanny, and D.~Lacoste.
\newblock {Role of ATP hydrolysis in the Dynamics of a single actin filament}.
\newblock {\em Biophys. J.}, 98:1418--1427, 2010.

\bibitem{mitchison:1984}
Tim Mitchison and Marc Kirschner.
\newblock Dynamic instability of microtubule growth.
\newblock {\em Nature}, 312:237--242, 1984.

\bibitem{sanoop2013}
Sanoop Ramachandran and Jean-Paul Ryckaert.
\newblock Compressive force generation by a bundle of living biofilaments.
\newblock {\em J. Chem. Phys.}, 139(6):--, 2013.

\bibitem{RobPhillips-book}
Rob Phillips, Jane Kondev, and Julie Theriot.
\newblock {\em Physical biology of the cell}.
\newblock Garland Science, New York, 2008.

\bibitem{Ingbar:03}
D.~E. Ingber.
\newblock Tensegrity i. cell structure and hierarchical systems biology.
\newblock {\em J. Cell. Sci.}, 116:1157--1173, 2003.

\bibitem{das-arxiv:2014}
Dipjyoti Das, Dibyendu Das, and Ranjith Padinhateeri.
\newblock Force-induced dynamical properties of multiple cytoskeletal filaments
  are distinct from that of single filaments.
\newblock {\em arXiv:1403.7708}, 2014.

\bibitem{Leibler:93}
Marileen Dogterom and Stanislas Leibler.
\newblock Physical aspects of the growth and regulation of microtubule
  structures.
\newblock {\em Phys. Rev. Lett.}, 70(9):1347--1350, Mar 1993.

\bibitem{jason-dogterom:03}
Marcel~E. Janson, Mathilde~E. de~Dood, and Marileen Dogterom.
\newblock {Dynamic instability of microtubules is regulated by force}.
\newblock {\em J. Cell Biol.}, 161(6):1029--1034, 2003.

\bibitem{Drechsel:1992}
DN~Drechsel, AA~Hyman, MH~Cobb, and MW~Kirschner.
\newblock Modulation of the dynamic instability of tubulin assembly by the
  microtubule-associated protein tau.
\newblock {\em Mol. Biol. Cell}, 3(10):1141--1154, 1992.

\bibitem{kolomeisky:04}
E.~B. Stukalin and A.~B. Kolomeisky.
\newblock Simple growth models of rigid multifilament biopolymers.
\newblock {\em J. Chem. Phys.}, 121:1097--1104, 2004.

\end{thebibliography}


\end{document}